\documentclass[
 reprint,
 amsmath,amssymb,
 superscriptaddress,
 aps,
 prl,
 nofootinbib,
]{revtex4-2}

\usepackage{siunitx}
\usepackage{graphicx}
\usepackage{dcolumn}
\usepackage{bm}
\usepackage[normalem]{ulem}
\usepackage{physics2}
\usephysicsmodule{ab}
\usephysicsmodule{op.legacy}
\usephysicsmodule{braket}
\usephysicsmodule{nabla.legacy}
\usepackage{derivative}
\usepackage{mathtools}

\usepackage{appendix}
\usepackage[colorlinks,linkcolor=blue,urlcolor=blue, citecolor=blue]{hyperref}
\usepackage[whole]{bxcjkjatype}
\usepackage[T1]{fontenc}
\usepackage{comment}
\usepackage{amsfonts}

\newcommand{\mqty}[1]{\ab(\begin{matrix}#1\end{matrix})}
\newcommand{\dd}[1]{\odif{#1}\,}

\usepackage{xcolor}

\usepackage{newtxtext}
\usepackage{newtxmath}
\begin{document}
\title{Scattering theory for cavity-assisted spin-motion-photon interactions}
\author{Seigo Kikura}
\email{seigo.kikura@nano-qt.com}
\affiliation{Nanofiber Quantum Technologies, Inc. (NanoQT), 1-22-3 Nishiwaseda, Shinjuku-ku, Tokyo 169-0051, Japan}
\author{Aruku Senoo}
\affiliation{Nanofiber Quantum Technologies, Inc. (NanoQT), 1-22-3 Nishiwaseda, Shinjuku-ku, Tokyo 169-0051, Japan}
\author{Akihisa Goban}
\email{akihisa.goban@nano-qt.com}
\affiliation{Nanofiber Quantum Technologies, Inc. (NanoQT), 1-22-3 Nishiwaseda, Shinjuku-ku, Tokyo 169-0051, Japan}
\author{Shinichi Sunami}
\email{shinichi.sunami@nano-qt.com}
\affiliation{Nanofiber Quantum Technologies, Inc. (NanoQT), 1-22-3 Nishiwaseda, Shinjuku-ku, Tokyo 169-0051, Japan}
\affiliation{Clarendon Laboratory, University of Oxford, Oxford OX1 3PU, United Kingdom}

\begin{abstract}
Cavity-assisted photon scattering (CAPS) is a powerful mechanism for realizing strong interactions between the internal states of stationary qubits and flying photons, underpinning a broad range of hybrid atom-photon protocols including remote entanglement generation and heralded atom-photon gates.
Recently, the motional quantum state has emerged as an important building block for quantum information processing with atomic qubits, both as a coherently controllable degree of freedom and as a fundamental error channel through undesired spin-motion coupling.
For the resonant-coupling regime of cavity quantum electrodynamics relevant to CAPS operations, however, the analytical formulation of spin-motion-photon coupling has so far remained elusive.
Here, we develop a complete analytical framework for CAPS that incorporates the coherent interaction between atomic motion and a reflected photon by extending scattering theory to include the motional degree of freedom.
The resulting compact operator-based input-output relation applies uniformly across various cavity geometries, spin-dependent trapping potentials, and nonidentical multiple spins.
As an exemplary application, we use the framework to elucidate how atomic motion affects CAPS-based atom-photon gates, identifying the parameter regimes that suppress motion-induced errors.
Our framework provides a theoretical foundation both for mitigating motional errors in CAPS operations and for deliberately exploiting motion-photon interaction at the atom-photon interface.
\end{abstract}
\maketitle

\emph{Introduction---}
Hybrid quantum systems that interface internal states of stationary qubits, such as atoms and ions, with flying photonic qubits underpin a broad range of quantum technologies, including distributed quantum computing~\cite{Monroe2014, Covey2023, Sunami2025, Main2024}, secure communication~\cite{Wehner2018, Azuma2023}, distributed sensing~\cite{Gottesman2012, Khabiboulline2019}, and blind quantum computing~\cite{Fitzsimons2017, Wei2025}.
In a free-space setting, however, the efficiency of stationary-flying-qubit interfaces is limited by the small overlap of the atomic dipole with the photon collection mode~\cite{Stephenson2020, Saha2025, Main2024}.
Optical cavities resolve these constraints and provide an ideal platform to interface atomic and photonic quantum information, enabling efficient remote entanglement generation~\cite{Ritter2012, Reiserer2015, Krutyanskiy2023, Hartung2024,Kikura2025_caps, Ji2026}, atom-light gates for discrete-variable photonic qubits~\cite{Duan2004, Duan2005, Lin2006, Volz2014, Daiss2021, Welte2021} and continuous-variable qumodes~\cite{Wang2005, Dhara2025, Kikura2026}, and the generation of nonclassical light based on such hybrid gates~\cite{Hacker2019, Hastrup2022}.

In parallel, atomic-qubit control has rapidly extended from the internal-state control to the motional-state control.
For trapped ions, motion-mediated entangling gates have reached the four-nines fidelity level~\cite{Loschnauer2025}, and bosonic codes such as the Gottesman--Kitaev--Preskill qubit have been demonstrated in the ion motion with state-of-the-art coherent control~\cite{Fluhmann2019, deNeeve2022, Matsos2025}.
For tweezer-trapped neutral atoms, recently, similar techniques have been developed to harness motion as a quantum resource, including coherent motion-spin control~\cite{Shaw2025, Tsai2026}, as well as motional state squeezing~\cite{Lienhard2025}.
Despite this progress, the role of atomic motion in atom-photon interaction protocols has only recently come into focus: in emission-based remote entanglement generation, the recoil kick associated with directional photon emission has been identified as a fundamental source of spin-motion-photon entanglement~\cite{Saha2025}, prompting analytical frameworks that quantify and mitigate the resulting motion-induced infidelity~\cite{Kikura2025_recoil, Apolin2025, Yu2025}.
In contrast, for the cavity-assisted photon scattering (CAPS) protocol for atom-photon and remote atom-atom gates~\cite{Duan2004, Duan2005, Lin2006}, a comparable description of motional effects has so far remained out of reach;
the analytical treatment has been restricted to the dispersive-coupling regime~\cite{Neumeier2018_PRA, Neumeier2018_IOP, Arguello2022}, which excludes the resonant operating conditions required for heralded CAPS atom-photon gates~\cite{Duan2004, Duan2005, Lin2006, Volz2014, Daiss2021, Welte2021,Kikura2025_caps}.

In this work, we close this gap by developing a complete analytical framework for CAPS that incorporates the coherent spin-motion-photon interaction induced by single-photon reflection from a cavity coupled to trapped atoms.
The key methodological step is to extend scattering theory~\cite{Fan2010, Shi2015} to include the motional dynamics, yielding a compact, operator-valued input-output relation that remains analytically tractable in the resonant-coupling regime relevant to high-fidelity CAPS gates.
The resulting framework applies uniformly across a broad range of settings, including standing- and running-wave cavities, spin-dependent trapping potentials, and multiple nonidentical spins coupled to a single cavity mode, and casts spin-motion-photon gate operations into a closed-form expression that can be evaluated efficiently.
As an exemplary demonstration of this framework, we evaluate how motional effects affect the performance of CAPS, such as the atom-photon controlled-phase gate, including realistic cases where the excited- and ground-state trap frequencies are different.
Beyond error analysis, by rendering the spin-motion-photon coupling explicit and analytically accessible, our framework opens a pathway to exploit it deliberately as a resource for hybrid quantum protocols at the atom-photon interface~\cite{Chakraborty2026,kemper2025}.

\begin{figure}[t]
    \centering
    \includegraphics[width=\linewidth]{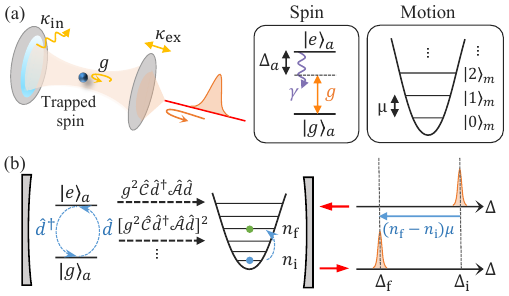}
    \caption{Concept of scattering theory for spin-motion-photon interactions.
    (a) System setup. A two-level system consisting of ground and excited states, $\ket{g}_a,\ket{e}_a$, is coupled to the cavity field at strength $g$.
    The detuning $\Delta_a = \omega_a-\omega_c$ between the spin-transition frequency $\omega_a$ and the cavity frequency $\omega_c$ is tuned for the desired operation.
    The cavity field couples with the output field through one of the cavity mirrors at rate $\kappa_\text{ex}$, while the cavity has intrinsic optical loss at rate $\kappa_\text{in}$; the total cavity-decay rate is $\kappa = \kappa_\text{ex} + \kappa_\text{in}$.
    The atom decays from the excited level at rate $\gamma$.
    For a trapped spin system with mass $m$ in a harmonic potential at frequency $\mu$, the motional state is described by the $n$-phonon basis $\ket{n}_m$.
    (b) Schematic of the scattering process. One cycle of the cyclic transition $\ket{g}_a\ket{1}_c \leftrightarrow \ket{e}_a\ket{0}_c$ incurs the action $g^2 \hat{\mathcal{C}}\hat{d}^\dagger \hat{\mathcal{A}}\hat{d}$ on the motional state. The phonon changes from $n_\text{i}$ to $n_\text{f}$ via photon reflection, where the reflected-photon frequency is shifted by $-(n_\text{f}-n_\text{i}) \mu$.
    }
    \label{fig1}
\end{figure}
\emph{CAPS-induced motion-photon interaction}---We first introduce the conventional two-level-system model before extending it to the model with a motional state.
Consider a motionally fixed spin consisting of ground and excited states $\ket{g}_a,\ket{e}_a$ at a position of maximum cavity electric field strength, resulting in a spin-photon coupling strength $g$ [see also Fig.~\ref{fig1}(a)].
Such fixed spin systems are realized by color centers in cavities~\cite{Knaut2024}, for example.
In the rotating frame of the cavity frequency $\omega_c$, the Hamiltonian of the atom-cavity system is given by $\hat{H}_\text{fixed} = \Delta_a \ketbra{e}{e}_a + g(\hat{c}\hat{\sigma}^\dagger + \text{h.c.})$, where $\Delta_a = \omega_a-\omega_c$ is the detuning of the atomic transition, $\hat{c}$ is the annihilation operator of the cavity mode, and $\hat{\sigma} = \ketbra{g}{e}_a$ ($\hbar = 1$ unless otherwise noted).
Here, we present the reflection function with conventional scattering theory, which is extended to incorporate atomic motion later.
In the interaction picture of the bath Hamiltonian $\hat{H}_B = \int \Delta \hat{a}^\dagger(\Delta)\hat{a}(\Delta) \odif{\Delta}$ where $\hat{a}(\Delta)$ is the annihilation operator of the propagating mode at the detuning $\Delta = \omega-\omega_c$, the scattering trajectory in the absence of atomic decay and cavity internal loss yields a pure spin-photon interaction.
This is characterized by an $S$ matrix for input and output frequencies, $\Delta_\text{i}$ and $\Delta_\text{f}$, as shown in~\cite{Fan2010}
\begin{equation} \label{eq:S_matrix_of_fixed_spin}
    \mathcal{S}_{\Delta_\text{f},\Delta_\text{i}} = \delta(\Delta_\text{f}-\Delta_\text{i}) r(\Delta_\text{i}),
\end{equation}
where
\begin{equation} \label{eq:definition_of_C_A}
    \begin{aligned}
        r(z) =& 1- 2i\kappa_\text{ex}\sum_{k=0}^{\infty}[g^2\mathcal{C}(z)\mathcal{A}(z)]^k\mathcal{C}(z), \\
        \mathcal{C}(z) \coloneqq& {}_a\bra{g}_c\bra{1} \hat{G}_0(z) \ket{g}_a\ket{1}_c = (z+i\kappa)^{-1}, \\
        \mathcal{A}(z) \coloneqq& {}_a\bra{e}_c\bra{0} \hat{G}_0(z) \ket{e}_a\ket{0}_c = (z-\Delta_a + i\gamma)^{-1}.
    \end{aligned}
\end{equation}
Here, $\ket{n}_c$ represents the $n$-photon state in the cavity mode, and we have introduced $\hat{G}_0(z) = 1/(z-\hat{\mathcal{H}}_0)$, which is the resolvent of $\hat{\mathcal{H}}_0 = (\Delta_a-i\gamma)\ketbra{e}[_a]{e} - i\kappa \hat{c}^\dagger\hat{c}$.
The term $g^2\mathcal{C}(z)\mathcal{A}(z)$ on the right-hand side of $r(z)$ represents the response of the cyclic transition $\ket{g}_a\ket{1}_c \to \ket{e}_a\ket{0}_c\to \ket{g}_a\ket{1}_c$ stimulated by the incident photon.

We now extend the above simple model to the atomic qubit confined in a harmonic potential; the following results can be extended to the case of multiple atoms coupled to a cavity (see the details in End Matter).
The atom position and momentum are respectively denoted by the quadrature operators $\hat{q}$ and $\hat{p}$, where we assume a one-dimensional system for simplicity; yet, the following results can be extended to a three-dimensional system, as explained in Sec.~\ref{ap_sec:sc_theory}.
We first address the case where spins in both ground and excited states experience the same trapping potential with a trap frequency $\mu$,
and discuss a state-dependent potential later.
The system energy is now added with the harmonic-potential term $\hat{H}_\text{ho} = \mu \hat{n}_b$, where $\hat{b} = (m\mu \hat{q} + i\hat{p})/\sqrt{2\hbar m\mu}$ is the phonon annihilation operator, and $\hat{n}_b = \hat{b}^\dagger\hat{b}$ is the phonon number operator.
This results in
\begin{equation} \label{eq:definition_of_hat_C_A}
    \begin{aligned}
        \hat{\mathcal{C}}(z) =& [(z+i\kappa)\hat{I}_m -\mu \hat{n}_b]^{-1}, \\
        \hat{\mathcal{A}}(z) =& [(z- \Delta_a + i\gamma)\hat{I}_m -\mu \hat{n}_b]^{-1},
    \end{aligned}
\end{equation}
where $\hat{I}_m$ is the identity operator in the motion subspace.
In addition, the cyclic transition $\ket{g}_a\ket{1}_c \to \ket{e}_a\ket{0}_c \to \ket{g}_a\ket{1}_c$ driven by photon incidence causes the motion-photon interaction as $g^2 \hat{\mathcal{C}}(z)\hat{d}^\dagger\hat{\mathcal{A}}(z)\hat{d}$, where $\hat{d}$ represents the motion-cavity coupling of the transition $\ket{g}_a\ket{n+1}_c \to \ket{e}_a\ket{n}_c$, determined by the electric-field amplitude of the cavity experienced by the atom [see also Fig.~\ref{fig1}(b)].

To analyze the motion-photon interaction, $S$ matrix of Eq.~\eqref{eq:S_matrix_of_fixed_spin} can be extended to track the phonon change $n_\text{i}\to n_\text{f}$ (see Sec.~\ref{ap_sec:sc_theory} for the detailed derivation),
\begin{equation} \label{eq:S^n_f,n_i,Delta_f,Delta_i}
    \mathcal{S}_{\Delta_\text{f},\Delta_\text{i}}^{n_\text{f},n_\text{i}} = \delta(\Delta_\text{f}-\Delta_\text{i}+(n_\text{f}-n_\text{i})\mu) \ _m\bra{n_\text{f}} \hat{\mathcal{S}}(\Delta_\text{i}) \ket{n_\text{i}}_m,
\end{equation}
where
\begin{equation}
    \hat{\mathcal{S}}(z) = \hat{I}_m - \frac{2\kappa_\text{ex}}{\kappa-i z} \sum_{n=0}^{\infty} \hat{\mathcal{R}}(z + n\mu) \ketbra{n}{n}_m,
\end{equation}
and $\ket{n}_m$ represents the $n$-phonon state.
Here, the delta function $\delta(\Delta_\text{f}-\Delta_\text{i}+(n_\text{f}-n_\text{i})\mu)$ in Eq.~\eqref{eq:S^n_f,n_i,Delta_f,Delta_i} represents the energy conservation; the phonon-number change $n_\text{f}-n_\text{i}$ shifts the photon frequency by $-(n_\text{f}-n_\text{i})\mu$.
Then, the operator $\hat{\mathcal{S}}(z)$, which is the extension of $r(z)$ in Eq.~\eqref{eq:definition_of_C_A}, includes the motion-photon interaction characterized by
\begin{equation}
    \hat{\mathcal{R}}(z) = [\hat{I}_m - g^2\hat{\mathcal{C}}(z)\hat{d}^\dagger\hat{\mathcal{A}}(z)\hat{d}]^{-1}.
\end{equation}
Thus, the reflected photon is no longer separable from the motional state, similar to photon emission cases~\cite{Kikura2025_recoil, Apolin2025, Yu2025}.

This scattering matrix is further translated to the operator that maps the initial motion-photon state to the postreflection one as
\begin{equation}
    \begin{aligned}
        &\sum_{n_\text{f},n_\text{i}} \ket{n_\text{f}}_m \ab[\iint \odif{\Delta_\text{f}}\dd{\Delta_\text{i}} \ket{\Delta_\text{f}}_p\mathcal{S}_{\Delta_\text{f},\Delta_\text{i}}^{n_\text{f},n_\text{i}} {}_p\bra{\Delta_\text{i}}] {}_m\bra{n_\text{i}}, \\
        &= \int \dd{\Delta} \hat{\mathcal{K}}(\Delta) \ketbra{\Delta}{\Delta}_p,
    \end{aligned}
\end{equation}
where $\ket{\Delta}_p = \hat{a}^\dagger(\Delta)\ket{\emptyset}_p$, $\ket{\emptyset}_p$ is the vacuum state of the propagating mode, and $\hat{\mathcal{K}}(\Delta)$ describes the motion-photon interaction induced by photon reflection; the explicit form is given as
\begin{equation}
    \hat{\mathcal{K}}(\Delta) = \hat{\mathcal{D}}(-\mu \hat{n}_b)\hat{\mathcal{S}}(\Delta) \hat{\mathcal{D}}(\mu \hat{n}_b),
\end{equation}
where we have defined $\hat{O}(\hat{n}_b) = \sum_{n} \hat{O}(n)\ketbra{n}{n}_m$ for any operator $\hat{O}(z)$ and introduced a frequency shift operator in up to one photon subspace, $\hat{\mathcal{D}}(\omega) = \int \dd{\Delta} \hat{a}^\dagger(\Delta + \omega) \hat{a}(\Delta)$, which satisfies $\hat{\mathcal{D}}(\omega)\ket{\Delta}_p = \ket{\Delta+\omega}_p$~\cite{Fabre2020, Fabre2022}.
Thus, our extended scattering theory derives the complete characterization of the motion-photon interaction induced by photon reflection, with the concise operator-based framework.
To illustrate the wide applicability of our framework, we have shown that our results reproduce the cavity-frequency shift and linewidth broadening in the dispersive-coupling regime, consistent with those obtained in Refs.~\cite{Neumeier2018_PRA, Neumeier2018_IOP} (see Sec.~\ref{ap_sec:dispersive_coupling}).

Our framework elucidates the sufficient condition for a recoil-free operation,
\begin{equation} \label{eq:recoil_free_condition}
    \hat{d}^\dagger \hat{d} = \hat{I}_m, \quad \text{and} \quad  \hat{d}^\dagger\hat{n}_b \hat{d} = \hat{n}_b,
\end{equation}
meaning that the motion-photon coupling operator $\hat{d}$ is a unitary operator that commutes with $\hat{n}_b$; the condition in Eq.~\eqref{eq:recoil_free_condition} gives
\begin{equation}
    \sum_{n} \hat{\mathcal{R}}(z + n\mu) \ketbra{n}{n}_m = \frac{1}{1-g^2 \mathcal{C}(z) \mathcal{A}(z)} \hat{I}_m,
\end{equation}
leading to $\mathcal{S}_{\Delta_\text{f},\Delta_\text{i}}^{n_\text{f},n_\text{i}} = \delta_{n_\text{f},n_\text{i}} \mathcal{S}_{\Delta_\text{f},\Delta_\text{i}}$.
However, this is not satisfied in realistic cavity implementations.
As an illustrative example, consider a running-wave cavity $\hat{d}_\text{run} = e^{ik\hat{q}} = \hat{D}_b(i\eta)$, where $\hat{D}_b(\beta) = e^{\beta\hat{b}^\dagger- \beta^\ast\hat{b}}$ is the displacement operator in the motion subspace and $\eta= k\sqrt{\hbar/(2m\mu)}$ represents the Lamb-Dicke (LD) parameter ($k$ is the cavity-field wavenumber, and $m$ is the mass of the atom).
The operator $\hat{d}_\text{run}$ is unitary but does not commute with $\hat{n}_b$, leading to the relation
\begin{equation} \label{eq:d_runAd_run}
    \hat{d}_\text{run}^\dagger\hat{\mathcal{A}}(z)\hat{d}_\text{run} = \ab[(z-\Delta_a+i\gamma)\hat{I}_m - \frac{k\hat{p}}{m} - \mu_\text{R}-\mu\hat{n}_b]^{-1}.
\end{equation}
This shows that the Doppler and recoil effects shift the excited-state energy by $\hbar k\hat{p}/m$ and $\hbar\mu_\text{R} = (\hbar k)^2/2m$, respectively.
Since $\hat{p}$ does not commute with $\hat{n}_b$, the operator $\hat{\mathcal{S}}(z)$ is not diagonal in the phonon number basis, clearly demonstrating the presence of a fundamental motion-photon-interaction effect in cavity photon scattering, which we analyze in more detail below.

\begin{figure}[t]
    \centering
    \includegraphics[width=\linewidth]{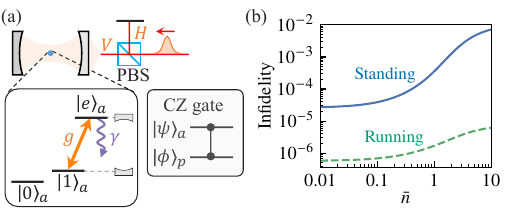}
    \caption{Atom-photon two-qubit gates with motion-photon coupling.
    (a) Schematic of CAPS-based two-qubit gates.
    The photonic qubit is encoded by its polarization, $\ket{\phi}_p = \alpha \ket{H}_p + \beta\ket{V}_p$, where $\ket{H(V)}_p$ represents the $H(V)$-polarized photon.
    Through a polarizing beamsplitter (PBS), only the $V$-polarized component is reflected by the cavity, resulting in a controlled-phase (CZ) gate between the atomic and photonic qubits (see the details in Ref.~\cite{Duan2004}).
    (b) Infidelity of a heralded phase gate.
    Here, the photon is prepared in $\ket{V}_p$ and detected, which translates the CZ gate into an atomic $Z$ gate.
    We employ the system parameters $(g,\kappa_\text{in},\gamma)/2\pi = (7.6, 0.2, 3)\, \unit{\MHz}$ and $\mu/2\pi = \qty{138}{\kHz}$ in Ref.~\cite{Seubert2025} that realized a $^{87}\text{Rb}$ atom coupled to a standing-wave cavity with the D2 line, which leads to the LD parameter $\eta = 0.17$.
    The solid and dashed lines respectively show the infidelity of the standing- and running-wave cavities.}
    \label{fig2}
\end{figure}
\emph{Motion effect in spin-photon two-qubit gates}---Motivated by the recent analysis of substantial motion effect in emission-based remote entanglement generation~\cite{Saha2025,Kikura2025_recoil,Apolin2025, Yu2025}, we now apply scattering theory to the analysis of motion-induced infidelity in the atom-photon controlled-phase (CZ) gate~\cite{Duan2004, Duan2005, Lin2006}.
Because conventional CAPS theory neglects atomic motion, the quantitative impact of motional effects on this canonical operation has so far remained unexplored, which is a crucial prerequisite for high-fidelity spin-photon CAPS operations.
Our framework provides the first analytical treatment of this regime and exposes the structure of the motion-photon interaction; furthermore, this has a clear structure, which will be turned into a useful protocol in future motion-photon hybrid protocols.

Here, instead of the two-level system, we consider the three-level system consisting of two stable (qubit) states, $\ket{0}_a$ and $\ket{1}_a$, and one excited state $\ket{e}_a$, where the transition $\ket{1}_a\leftrightarrow\ket{e}_a$ resonantly couples to the cavity $(\Delta_a = 0)$, as shown in Fig.~\ref{fig2}(a).
For the atom being initially prepared in qubit states $\{\ket{0}_a,\ket{1}_a\}$, the photon reflection gives the state-dependent $\pi$ phase shift, enabling atom-photon two-qubit gates acting on the polarization-encoded photonic qubit with linear optics~\cite{Duan2004}.

To evaluate the worst-case scenario in terms of motion-induced errors, we initialize the photon in the $V$-polarized state, which directly couples to the cavity;
furthermore, to see the effect on the heralded gate, we assume that the reflected photon is subsequently detected by a photon detector; this maps photon-loss effects onto the success probability of the two-qubit gate.
In the following, we set the output coupling rate as $\kappa_\text{ex} = \kappa_\text{in}\sqrt{1+2C_\text{in}}$, where $C_\text{in} = g^2/2\kappa_\text{in}\gamma$ is the internal cooperativity, to maximize gate fidelity in the absence of the motion-induced error~\cite{Goto2010}.
Under this configuration, the gate infidelity arises exclusively from the spin-motion-photon interaction.
For the atom prepared in $\ket{+}_a = (\ket{0}_a + \ket{1}_a)/\sqrt{2}$, we evaluate the infidelity with the target state $\ket{-}_a = (\ket{0}_a - \ket{1}_a)/\sqrt{2}$ in the standing- and running-wave cavities, as shown in Fig.~\ref{fig2}(b).

Because the standing-wave cavity coherently induces two opposite kicks,
\begin{equation}
    \hat{d}_\text{st} = \cos(k\hat{q}) = \frac{\hat{D}_b(i\eta) + \hat{D}_b(-i\eta)}{2},
\end{equation}
the cyclic transition $\ket{g}_a\ket{1}_c \leftrightarrow \ket{e}_a\ket{0}_c $ induces recoil kicks at amplitude $\pm 2i\eta$, resulting in higher infidelity than the running-wave cavity, in which recoil kicks do not occur twice in the same direction.
In addition, the standing-wave cavity exhibits a steeper increase in infidelity as a function of the mean phonon number $\bar{n}$ of the prepared thermal motional state.
For the motional carrier transition, the motion-cavity coupling can be translated into the effective coupling strength operator (see the details in Sec.~\ref{ap_sec:phase_inversion})
\begin{equation}
    \hat{g}_\text{st} = \ab[1-\ab(\hat{n}_b + \frac{1}{2})\eta^2]g,
\end{equation}
as in photon-emission cases~\cite{Kikura2025_recoil}.
This means that different phonon-number states reflect the photon with different reflection amplitudes, leading to a stronger dependence on $\bar{n}$.
In contrast, for a running-wave cavity, the effect is captured by the Doppler and recoil shifts in Eq.~\eqref{eq:d_runAd_run} with weaker dependence on the phonon number.

To compare with the photon-emission cases~\cite{Kikura2025_recoil}, we further investigate the dependence on the parameter ratios $\kappa/\mu$ and $\kappa/g$ (see Sec.~\ref{ap_sec:phase_inversion}).
The dependence on $\kappa/\mu$ shows behavior similar to the emission case: the sideband-resolved regime, $\kappa<\mu$, exhibits lower infidelity than the unresolved one, $\kappa>\mu$.
In contrast, a wide range of $\kappa/g$ exhibits nearly constant infidelity, while the emission case shows lower infidelity with $\kappa>g$.
This indicates that CAPS-based gates are robust to motion-induced effects, independent of the cavity-QED regimes, including the strong-coupling regime ($g>\kappa,\gamma$) or the bad-cavity regime ($\kappa>g>\gamma$), provided that the output coupling rate $\kappa_\text{ex}$ is optimized accordingly.\footnote{We note that the comparison with the results of Ref.~\cite{Kikura2025_recoil} can be made by increasing the reported infidelities by a factor of two; see also the Supplemental Material.}

\begin{figure}[t]
    \centering
    \includegraphics[width=\linewidth]{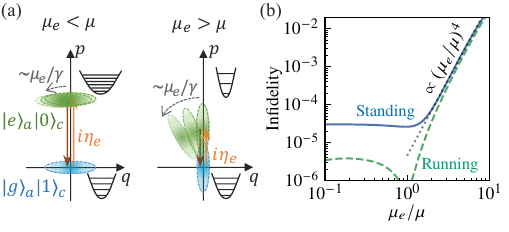}
    \caption{Effect arising from state-dependent trapping potentials.
    (a) Conceptual description of motional-state dynamics via one cyclic transition $\ket{g}_a\ket{1}_c \to \ket{e}_a\ket{0}_c\to \ket{g}_a\ket{1}_c$ for a running-wave cavity. The motional state is first squeezed by $\hat{S}_b(\zeta)$ before moving to the excited-state potential, followed by displacement by amplitude $i\eta_e$.
    The state is then distorted by phase rotation at speed $\mu_e$ within the duration characterized by the lifetime $1/\gamma$ (captured by $\hat{\mathcal{A}}$). The state finally comes back to the ground-state potential by $\hat{S}_b^\dagger(\zeta)\hat{D}_b(-i\eta_e)$.
    Since a $q$-squeezed state is more sensitive to the phase rotation, and the rotation-angle rate is the excited-state trap frequency, the case of $\mu_e > \mu$ disturbs the motional state more than that of $\mu_e < \mu$.
    (b) Infidelity induced by the deviation from the magic trap condition, $\mu_e=\mu$, where the initial motional state is the vacuum state.
    For $\mu_e > \mu$, the infidelity scales approximately as $(\mu_e/\mu)^4$, indicated by the dotted line, in both cavities.}
    \label{fig3}
\end{figure}
\emph{State-dependent trap frequencies}---We now consider the case where the excited state experiences a trap frequency $\mu_e$ different from that of the ground state, $\mu$; the harmonic-potential term is replaced with~\cite{Taieb1994}
\begin{equation}
    \hat{H}_\text{ho} = \ketbra{g}{g}_a \otimes \mu \hat{b}^\dagger\hat{b} + \ketbra{e}{e}_a \otimes \ab(\mu_e\hat{b}_e^\dagger\hat{b}_e + \frac{\mu_e-\mu}{2}),
\end{equation}
where the excited-state annihilation operator is defined as $\hat{b}_e = (m\mu_e \hat{q} + i\hat{p})/\sqrt{2\hbar m\mu_e}$.
Because the excited-state trapping potential enters the scattering operator only through the excited-state resolvent $\hat{\mathcal{A}}(z)$ in Eq.~\eqref{eq:definition_of_hat_C_A}, the inhomogeneity of the trap frequencies is incorporated simply by replacing its effective energy:
\begin{equation}
    (\Delta_a - i\gamma)\hat{I}_m +\mu \hat{n}_b \to \ab(\Delta_a - i\gamma + \frac{\mu_e -\mu}{2})\hat{I}_m +\mu_e \hat{b}_e^\dagger \hat{b}_e.
\end{equation}
Furthermore, from the relation $\hat{b}_e = \hat{S}_b^\dagger(\zeta)\hat{b}\hat{S}_b(\zeta)$, where $\hat{S}_b(r) = e^{r/2(\hat{b}^2-\hat{b}^{\dagger 2})}$ is the squeezing operator in the motion subspace, and $2\zeta = -\ln(\mu_e/\mu)$~\cite{Taieb1994}, we find that the framework in the same trap frequency, $\mu_e = \mu$, remains valid by replacing
\begin{equation}
    \hat{\mathcal{A}}(z) \to \hat{S}_b^\dagger(\zeta)\hat{\mathcal{A}}(z)\hat{S}_b(\zeta),
\end{equation}
along with the extension of $\hat{\mathcal{A}}(z)$ as
\begin{equation} \label{eq:extension_of_A}
    \hat{\mathcal{A}}(z) = \ab[(z- \Delta_a + i\gamma)\hat{I}_m -\mu \hat{n}_b - \Delta\mu \ab(\hat{n}_b + \frac{1}{2})]^{-1},
\end{equation}
where $\Delta\mu = \mu_e - \mu$.
By using the extended framework, we again evaluate the motion-induced infidelity in photon reflection.
From the relation $\hat{S}_b(\zeta)\hat{D}_b(\pm i\eta) = \hat{D}_b(\pm i\eta_e) \hat{S}_b(\zeta)$, where $\eta_e = k\sqrt{\hbar/(2m\mu_e)}$ represents the LD parameter of the excited state, the motion-state action of the transition $\ket{g}_a \rightarrow \ket{e}_a$ can be interpreted as the squeezing followed by the displacement by amplitude $i\eta_e$.
Then, the motional state is distorted by $\hat{\mathcal{A}}(z)$, which includes the phase rotation $\sim e^{-i\mu_e \hat{n}_b/\gamma}$, followed by $\hat{S}_b^\dagger(\zeta)\hat{D}_b(-i\eta_e)$, as shown in Fig.~\ref{fig3}(a) (see the detailed discussion in Sec.~\ref{ap_sec:phase_inversion}).
Figure~\ref{fig3}(b) shows that for $\mu_e>\mu$ the infidelity in both running- and standing-wave cavities scales as $(\mu_e/\mu)^4$, since the squeezed state is more susceptible to the phase rotation induced by $\hat{\mathcal{A}}(z)$.
In contrast, for $\mu_e < \mu$, the infidelity does not increase in such an unfavorable scaling; a large deviation from the magic trap condition ($\mu_e=\mu$)  suppresses the infidelity.
Overall, our framework provides useful insight for experiments employing trapped atoms coupled to optical cavities:
finite atomic temperature effects and motion-induced errors are below $10^{-3}$ level for CAPS atom-photon gates if the external coupling rate $\kappa_\mathrm{ex}$ is chosen appropriately, including in cases where the trap is not necessarily magic.

\emph{Conclusion}---We have developed a complete analytical description of CAPS that incorporates coherent spin-motion-photon interaction, previously overlooked in theories for CAPS.
By extending scattering theory to include motional degrees of freedom, we obtain a compact operator-based input-output relation applicable to diverse setups; while we have focused on a single-spin system in the main text, our framework is also consistently applicable to the case of multiple spins coupled to a cavity (see the details in End Matter).
As exemplary applications, we have used the framework to analyze the impact of atomic motion on the atom-photon CZ gate, identifying parameter regimes in which motional effects are enhanced and suppressed.
Together with the recent development of highly multiplexed operations for CAPS-based remote entanglement generation~\cite{Kikura2025_caps}, our work further establishes CAPS as a robust means for high-rate, high-fidelity remote entanglement generation.
More broadly, by rendering the spin-motion-photon coupling explicit and analytically tractable, the present framework provides a unified theoretical foundation for deliberately exploiting motion-photon interaction as a resource for hybrid quantum applications at the atom-photon interface.

\emph{Acknowledgments}---S.K. and S.S. are employees, A.S. is an intern, and A.G. is a co-founder and a shareholder of Nanofiber Quantum Technologies, Inc.

\emph{Note added}---During the preparation of this manuscript, we became aware of a related work on levitated optomechanics~\cite{lednev2026engineeringrecoilheatingcoherentscattering}.

\clearpage
\begin{widetext}
    \centering
    \textbf{End Matter}
\end{widetext}

\section{Multi-atom case}
In the main text, we focused on the case where a cavity hosts only a two-level atom.
In contrast, multiple atoms coupled to a single cavity significantly extend the functionality of CAPS, such as local multiple-qubit gates~\cite{Lin2006, Welte2018}, efficient preparation of entangled states shared by many atoms~\cite{Nagib2025}, as well as time- and wavelength-multiplexed operations~\cite{Kikura2025_caps}.
As outlined below, the resulting many-body motion-photon interaction can be addressed by our developed scattering theory.
Here, $N$ atoms couple with a single cavity mode, where parameters and operators for the $j$th atom are labeled by the subscript $j \in \mathsf{Q} = \{1,2,\cdots,N\}$.
As in the single-atom case, we derive the operator that maps a pre-reflection state consisting of a photon and $N$ motional states to the post-reflection one as $\int \dd{\Delta} \hat{\mathcal{K}}^{(N)}(\Delta) \hat{a}^\dagger(\Delta)\hat{a}(\Delta)$ (see the detailed derivation in Sec.~\ref{ap_sec:sc_theory}), where
\begin{equation}
    \hat{\mathcal{K}}^{(N)}(\Delta) = \hat{\mathcal{D}}(-\bm{\mu}^\top \hat{\bm{n}}_b)\hat{\mathcal{S}}^{(N)}(\Delta) \hat{\mathcal{D}}(\bm{\mu}^\top \hat{\bm{n}}_b).
\end{equation}
Here, $\bm{n} = (n_1,n_2,\cdots, n_N)$ and $\bm{\mu} = (\mu_1,\mu_2,\cdots, \mu_N)$ are the vectors consisting of phonon numbers and trap frequencies, respectively.
Note that $\hat{O}(\bm{\mu}^\top \hat{\bm{n}}_b) = \sum_{\bm{n}} \hat{O}(\bm{\mu}^\top \bm{n}) (\bigotimes_{j\in \mathsf{Q}} \ketbra*{n_j}{n_j}_{m,j})$ for any operator $\hat{O}(z)$.
The $S$ operator is derived as
\begin{equation}
    \hat{\mathcal{S}}^{(N)}(\Delta) = \hat{I}_m^{(N)} - \frac{2\kappa_\text{ex}}{\kappa-i\Delta} \sum_{\bm{n}} \hat{\mathcal{R}}^{(N)}(\Delta + \bm{\mu}^\top \bm{n}) \ketbra{\bm{n}}{\bm{n}}_m,
\end{equation}
where $\hat{I}_m^{(N)}$ is the identity operator in the motion subspace, $\ket{\bm{n}}_m = \ket{n_1}_{m,1}\ket{n_2}_{m,2}\cdots\ket{n_N}_{m,N}$, and
\begin{equation}
    \begin{aligned}
        &\hat{R}^{(N)}(z) \\
        &= \ab[\hat{I}_m^{(N)}-\hat{\mathcal{C}}^{(N)}(z) \sum_{j\in \mathsf{Q}} g_j^2 \hat{d}_j^\dagger \hat{S}_{b,j}^\dagger(\zeta_j) \hat{\mathcal{A}}_j^{(N)}(z) \hat{S}_{b,j}(\zeta_j) \hat{d}_j]^{-1}.
    \end{aligned}
\end{equation}
Here, $\hat{\mathcal{C}}^{(N)}(z)$ and $\hat{\mathcal{A}}_j^{(N)}(z)$ are defined as the extension of $\hat{\mathcal{C}}(z)$ and $\hat{\mathcal{A}}(z)$ in Eqs.~\eqref{eq:definition_of_hat_C_A} and \eqref{eq:extension_of_A},
\begin{equation}
    \begin{aligned}
        \hat{\mathcal{C}}^{(N)}(z) =& \ab[(z+i\kappa)\hat{I}_m - \hat{H}_{\text{ho},g}^{(N)}]^{-1}, \\
        \hat{\mathcal{A}}_j^{(N)}(z) =& \ab[(z- \Delta_{a,j} + i\gamma_j)\hat{I}_m -\hat{H}_{\text{ho},g}^{(N)}- \Delta\mu_j\ab(\hat{n}_{b,j} + \frac{1}{2})]^{-1},
    \end{aligned}
\end{equation}
where we have defined the harmonic-potential term of all spins to be the ground states as
\begin{equation}
    \hat{H}_{\text{ho},g}^{(N)} = \sum_{k\in \mathsf{Q}} \mu_k\hat{n}_{b,k}.
\end{equation}
These results allow us to faithfully address fundamental yet previously unexplored motion-photon quantum many-body dynamics.
Applying our results to evaluate concrete motion-spin-photon effects in diverse scenarios is left for future work.

\clearpage

\begin{widetext}
\begin{center}
\textbf{\large Supplemental Material for ``Scattering theory for cavity-assisted spin-motion-photon interactions''}
\end{center}
\end{widetext}

\setcounter{equation}{0}
\renewcommand{\theequation}{S\arabic{equation}}
\setcounter{figure}{0}
\renewcommand{\thefigure}{S\arabic{figure}}
\newcounter{supplsec}
\renewcommand{\thesupplsec}{S\arabic{supplsec}}

\newcommand{\supsection}[1]{
  \refstepcounter{supplsec}
  \begin{center}
  \noindent\textbf{\thesupplsec: #1}\par
  \end{center}
}
\newcounter{supplsubsec}[supplsec]
\renewcommand{\thesupplsubsec}{\thesupplsec.\arabic{supplsubsec}}
\newcommand{\supsubsection}[1]{
  \refstepcounter{supplsubsec}
  \begin{center}
  \noindent\textbf{\thesupplsubsec\quad #1}\par
  \end{center}
}

\supsection{Extended scattering theory to incorporate motional states} \label{ap_sec:sc_theory}

Here, we extend scattering theory to address motion-photon interaction induced by photon reflection.
For readability, we first consider the case of a single two-level system coupled to a cavity in Sec.~\ref{ap_subsec:sc_theory_single_two-level_system}, and then apply our framework to a case of multiple two-level systems in Sec.~\ref{ap_subsec_N_two-level_system_case}.

\supsubsection{Case of a single two-level system} \label{ap_subsec:sc_theory_single_two-level_system}

To derive a scattering operator acting on the motional degree of freedom (DOF), we employ the scattering theory for quantum optics originally developed in Refs.~\cite{Shi2015, Caneva2015}.
The authors in Ref.~\cite{Shi2015} derived the reflection function $r(\Delta)$, in the absence of the two-level-system motion, as an exemplary result obtained by their developed theory.
Here, based on the alternative introduction of the scattering theory in Ref.~\cite{Chang2016}, we extend it to incorporate the motional DOF.

A cavity-QED system couples the propagating mode through the coupling mirror at rate $\kappa_\text{ex}$.
The total Hamiltonian, which includes system-decay terms, is given by (here, we explicitly present $\hbar$)
\begin{equation}
    \hat{H} = \hat{H}_\text{sys} + \hat{H}_\text{ho} + \hat{H}_B + \hat{H}_\text{int} + \hat{H}_\text{decay},
\end{equation}
where
\begin{equation}
    \begin{aligned}
        \hat{H}_\text{ho} =& \ketbra{g}{g}_a \otimes \hbar\mu \hat{b}^\dagger\hat{b} + \ketbra{e}{e}_a \otimes  \hbar \ab(\mu_e\hat{b}_e^\dagger\hat{b}_e + \frac{\mu_e-\mu}{2}),  \\
        \hat{H}_B =& \int \dd{\Delta} \hbar \Delta \hat{a}^\dagger(\Delta)\hat{a}(\Delta), \\
        \hat{H}_\text{int} =&  i\hbar\sqrt{\frac{\kappa_\text{ex}}{\pi}} \int \dd{\Delta} [\hat{a}^\dagger(\Delta) \hat{c}-\hat{c}^\dagger \hat{a}(\Delta)],
    \end{aligned}
\end{equation}
and $\Delta = \omega-\omega_c$ represents the detuning of the field from the cavity.
The decay term $\hat{H}_\text{decay}$ consists of the coupling of the excited state of the two-level system and the cavity mode with the environment, which imposes the spontaneous emission at rate $\gamma$ and the internal cavity loss at rate $\kappa_\text{in}$, respectively.
The local system Hamiltonian is given by
\begin{equation}
    \hat{H}_\text{sys} = \hbar\Delta_a \ketbra{e}{e}_a + \hat{V},
\end{equation}
where
\begin{equation}
    \begin{aligned}
        \hat{V} =& \hbar g(\hat{c}\hat{\sigma}^\dagger \hat{d} + \hat{c}^\dagger\hat{\sigma} \hat{d}^\dagger).
    \end{aligned}
\end{equation}
In the interaction picture with respect to $\hat{H}_\text{f} = \hat{H}_B + \hat{H}_\text{ho}$, the unitary propagator is given by
\begin{equation}\label{eq:propagator}
    \hat{U}_\text{I}(t_\text{f};t_\text{i}) = \hat{U}_\text{f}^\dagger(t_\text{f}) e^{-i\hat{H}(t_\text{f}-t_\text{i})/\hbar} \hat{U}_\text{f}(t_\text{i}),
\end{equation}
where
\begin{equation}
    \hat{U}_\text{f}(t) = e^{-i (\hat{H}_B +\hat{H}_\text{ho}) t/\hbar}.
\end{equation}
This gives the $S$ operator, which acts on the motion subspace, as~\cite{Newton1982, Shi2015, Xu2015}
\begin{equation} \label{ap_eq:S_operator}
    \hat{\mathcal{S}}_{\Delta_\text{f},\Delta_i} = \lim_{t_\text{f}\to \infty, t_\text{i}\to -\infty} \bra{\text{v}}\hat{a}(\Delta_\text{f}) \hat{U}_\text{I}(t_\text{f};t_\text{i})\hat{a}^\dagger(\Delta_\text{i})\ket{\text{v}},
\end{equation}
where $\ket{\text{v}}$ represents a state with no excitation of the two-level system, the cavity mode, the propagating mode, and the environment.
In the following, we may omit the limit notation for notational simplicity.
By using the functional derivatives
\begin{equation}
    \fdv{}{J_{\Delta}} e^{\int \dd{\Delta^\prime} J_{\Delta^\prime} \hat{a}^\dagger(\Delta^\prime)} = \hat{a}^\dagger(\Delta) e^{\int \dd{\Delta^\prime} J_{\Delta^\prime} \hat{a}^\dagger(\Delta^\prime)},
\end{equation}
we find
\begin{equation}
    \begin{aligned}
        &\hat{a}(\Delta_\text{f}) \hat{U}_\text{I}(t_\text{f};t_\text{i})\hat{a}^\dagger(\Delta_\text{i})\\
        &= \fdv{}{J_{\Delta_\text{f}}^\ast} \fdv{}{J_{\Delta_\text{i}}} e^{\int \dd{\Delta} J_{\Delta}^\ast \hat{a}(\Delta)} \hat{U}_\text{I}(t_\text{f};t_\text{i}) e^{\int \dd{\Delta} J_{\Delta} \hat{a}^\dagger(\Delta)}\Big|_{\{J_{\Delta}, J_{\Delta}^\ast\}\to 0},
    \end{aligned}
\end{equation}
where $\{J_\Delta\}$ are treated as independent complex variables for each frequency mode $\hat{a}(\Delta)$.
By introducing a displacement operator
\begin{equation}
    \begin{aligned}
        \hat{D}(\{J_\Delta\}) &= e^{\int \dd{\Delta} [J_\Delta \hat{a}^\dagger(\Delta) - J_\Delta^\ast \hat{a}(\Delta)]} \\
        &= e^{\int \dd{\Delta} J_\Delta \hat{a}^\dagger(\Delta)} e^{\int \dd{\Delta} [-J_\Delta^\ast \hat{a}(\Delta)]} e^{-\int \dd{\Delta} |J_\Delta|^2/2},
    \end{aligned}
\end{equation}
we obtain
\begin{equation} \label{ap_eq:<v|eUe|v>}
    \begin{aligned}
        &\bra{\text{v}}e^{\int \dd{\Delta} J_{\Delta}^\ast \hat{a}(\Delta)} \hat{U}_\text{I}(t_\text{f};t_\text{i}) e^{\int \dd{\Delta} J_{\Delta} \hat{a}^\dagger(\Delta)}\ket{\text{v}} \\
        &= \bra{\text{v}}\hat{U}_D(t_\text{f};t_\text{i}) \ket{\text{v}} e^{\int \dd{\Delta} |J_\Delta|^2},
    \end{aligned}
\end{equation}
where
\begin{equation} \label{ap_eq:def_U_D}
    \hat{U}_D(t_\text{f};t_\text{i}) = \hat{D}^\dagger(\{J_\Delta\}) \hat{U}_\text{I}(t_\text{f};t_\text{i}) \hat{D}(\{J_\Delta\}).
\end{equation}
From the relation
\begin{equation}
    \fdv{}{J_{\Delta_\text{f}}^\ast} \fdv{}{J_{\Delta_\text{i}}}e^{\int \dd{\Delta} |J_\Delta|^2}\Big|_{\{J_{\Delta}, J_{\Delta}^\ast\}\to 0} = \delta(\Delta_\text{f}-\Delta_\text{i}),
\end{equation}
and Eqs.~\eqref{ap_eq:S_operator} and~\eqref{ap_eq:<v|eUe|v>}, we then derive
\begin{equation} \label{ap_eq:S_Delta_f_Delta_i}
    \begin{aligned}
        &\hat{\mathcal{S}}_{\Delta_\text{f},\Delta_i} \\
        &= \fdv{}{J_{\Delta_\text{f}}^\ast} \fdv{}{J_{\Delta_\text{i}}} \bra{\text{v}}\hat{U}_D(t_\text{f};t_\text{i}) \ket{\text{v}} e^{\int \dd{\Delta} |J_\Delta|^2}\Big|_{\{J_{\Delta}, J_{\Delta}^\ast\}\to 0} \\
        &= \delta(\Delta_\text{f}-\Delta_\text{i}) \hat{I}_m  \\
        & \hspace{0.5cm} +e^{\int \dd{\Delta} |J_\Delta|^2}\fdv{}{J_{\Delta_\text{f}}^\ast} \fdv{}{J_{\Delta_\text{i}}} \bra{\text{v}}\hat{U}_D(t_\text{f};t_\text{i})\ket{\text{v}} \Big|_{\{J_{\Delta}, J_{\Delta}^\ast\}\to 0},
    \end{aligned}
\end{equation}
since $\hat{U}_\text{I}(t_\text{f};t_\text{i})\ket{\text{v}} = \ket{\text{v}}$.
The first term on the right-hand side of Eq.~\eqref{ap_eq:S_Delta_f_Delta_i} represents the dynamics where the input wave does not interact with the local system, while the second term corresponds to the scattering, which we further calculate in the following.

Here, we rewrite the unitary propagator as
\begin{equation}
    \hat{U}_\text{I}(t_\text{f};t_\text{i}) = \mathcal{T}\ab[e^{- \frac{i}{\hbar}\int_{t_\text{i}}^{t_\text{f}} \dd{t} \hat{H}_\text{I}(t) }],
\end{equation}
where $\mathcal{T}[\cdot]$ represents the time-ordering superoperator, and
\begin{equation}
    \hat{H}_\text{I}(t) = \hat{U}_\text{f}^\dagger(t) (\hat{H}- \hat{H}_\text{f}) \hat{U}_\text{f}(t).
\end{equation}
Since $\hat{H}_\text{decay}$ commutes with $\hat{H}_\text{f}$, it reduces to
\begin{equation}
    \hat{H}_\text{I}(t) = \hat{H}_\text{sys,I}(t) + \hat{H}_\text{int,I}(t) + \hat{H}_\text{decay},
\end{equation}
where
\begin{equation}
    \begin{aligned}
        \hat{H}_\text{sys,I}(t) =& \hat{U}_\text{f}^\dagger(t) \hat{H}_\text{sys} \hat{U}_\text{f}(t), \\
        \hat{H}_\text{int,I}(t) =& i\hbar\sqrt{\frac{\kappa_\text{ex}}{\pi}} \int \dd{\Delta} \ab[\hat{a}^\dagger(\Delta) \hat{c} e^{i\Delta t}-\hat{c}^\dagger \hat{a}(\Delta) e^{-i\Delta t}].
    \end{aligned}
\end{equation}
From the relation
\begin{equation}
    \hat{D}^\dagger(\{J_\Delta\}) \hat{H}_\text{int,I}(t) \hat{D}(\{J_\Delta\}) = \hat{H}_\text{int,I}(t) + \hat{H}_D(t),
\end{equation}
where
\begin{equation}
    \hat{H}_D(t) = i\hbar\sqrt{\frac{\kappa_\text{ex}}{\pi}} \int \dd{\Delta} \ab(J_{\Delta}^\ast \hat{c} e^{i\Delta t}-\hat{c}^\dagger J_{\Delta} e^{-i\Delta t}),
\end{equation}
Eq.~\eqref{ap_eq:def_U_D} gives
\begin{equation}
    \begin{aligned}
        &\hat{U}_D(t_\text{f};t_\text{i}) \\
        &= \mathcal{T}\ab[\exp\ab\{-\frac{i}{\hbar}\int_{t_\text{i}}^{t_\text{f}} \dd{t} [\hat{H}_\text{I}(t) + \hat{H}_D(t)]\}] \\
        &= \mathcal{T}\ab[\exp\ab\{-\frac{i}{\hbar}\int_{t_\text{i}}^{t_\text{f}} \dd{t} \hat{U}_\text{f}^\dagger(t) \ab[(\hat{H}-\hat{H}_\text{f}) + \hat{H}_D(t)] \hat{U}_\text{f}(t)\}].
    \end{aligned}
\end{equation}
By using the following relation
\begin{equation}
    \begin{aligned}
        &\mathcal{T}\Bigg[\exp \ab[-i \int_{t_0}^{t_1}\dd{t}e^{i\hat{O}_0 t}\hat{O}_1(t)e^{-i\hat{O}_0 t}] \Bigg]  \\
        &=  e^{i\hat{O}_0 t_1} \mathcal{T}\ab[e^{-i \int_{t_0}^{t_1}\dd{t} [\hat{O}_0 + \hat{O}_1(t)] }] e^{-i\hat{O}_0 t_0},
    \end{aligned}
\end{equation}
for any operators $\hat{O}_0$ and $\hat{O}_1(t)$, we derive
\begin{equation}
    \begin{aligned}
        &\hat{U}_D(t_\text{f};t_\text{i}) \\
        &= e^{i \hat{H}_\text{f} t_\text{f}/\hbar}\mathcal{T}\ab[e^{- \frac{i}{\hbar}\int_{t_\text{i}}^{t_\text{f}}\dd{t} [\hat{H} + \hat{H}_D(t)]}] e^{-i\hat{H}_\text{f} t_\text{i}/\hbar} \\
        &= e^{i (\hat{H}_\text{f} - \hat{H}) t_\text{f}/\hbar}\mathcal{T}\ab[e^{-\frac{i}{\hbar}\int_{t_\text{i}}^{t_\text{f}}\dd{t} \hat{H}_{D,\text{H}}(t)}]e^{-i(\hat{H}_\text{f}-\hat{H}) t_\text{i}/\hbar},
    \end{aligned}
\end{equation}
where $\hat{H}_{D,\text{H}}(t)$ represents $\hat{H}_D(t)$ in the Heisenberg picture,
\begin{equation}
    \begin{aligned}
        \hat{H}_{D,\text{H}}(t) =&  e^{i\hat{H} t/\hbar}\hat{H}_D(t) e^{-i\hat{H} t/\hbar}\\
        =& i\hbar\sqrt{\frac{\kappa_\text{ex}}{\pi}} \int \dd{\Delta} [J_{\Delta}^\ast \hat{c}_\text{H}(t) e^{i\Delta t}-\hat{c}_\text{H}^\dagger(t) J_{\Delta} e^{-i\Delta t}],
    \end{aligned}
\end{equation}
and $\hat{c}_\text{H}(t) = e^{i\hat{H}t/\hbar} \hat{c}e^{-i\hat{H}t/\hbar}$.
From the relation
\begin{equation}
    (\hat{H}- \hat{H}_\text{f})\ket{\text{v}} = (\hat{H}_\text{sys}+ \hat{H}_\text{int} + \hat{H}_\text{decay})\ket{\text{v}} = 0,
\end{equation}
we find
\begin{equation}
    \begin{aligned}
        \bra{\text{v}}\hat{U}_D(t_\text{f};t_\text{i})\ket{\text{v}} = \bra{\text{v}} \mathcal{T}\ab[e^{-\frac{i}{\hbar}\int_{t_\text{i}}^{t_\text{f}}\dd{t} \hat{H}_{D,\text{H}}(t)}] \ket{\text{v}}.
    \end{aligned}
\end{equation}
From the Dyson expansion
\begin{equation}
    \begin{aligned}
        &\mathcal{T}\ab[e^{-i \int_{t_\text{i}}^{t_\text{f}}\dd{t} \hat{O}(t)}] \\
        &= 1- i\int_{t_\text{i}}^{t_\text{f}}\dd{t_1}\hat{O}(t_1) - \int_{t_\text{i}}^{t_\text{f}}\odif{t_1}\int_{t_\text{i}}^{t_1}\dd{t_2}\hat{O}(t_1)\hat{O}(t_2) + \cdots,
    \end{aligned}
\end{equation}
we further obtain
\begin{equation} \label{ap_eq:second_term_calc_1}
    \begin{aligned}
        &e^{\int \dd{\Delta} |J_\Delta|^2}\fdv{}{J_{\Delta_\text{f}}^\ast} \fdv{}{J_{\Delta_\text{i}}} \bra{\text{v}}\hat{U}_D(t_\text{f};t_\text{i})\ket{\text{v}} \Big|_{\{J_{\Delta}, J_{\Delta}^\ast\}\to0}  \\
        &= -\frac{\kappa_\text{ex}}{\pi} \int_{t_\text{i}}^{t_\text{f}} \odif{t_1} \int_{t_\text{i}}^{t_1} \dd{t_2} \bra{\text{v}}\hat{c}_\text{H}(t_1)\hat{c}_\text{H}^\dagger(t_2)\ket{\text{v}} e^{i(\Delta_\text{f}t_1 -\Delta_\text{i} t_2)}. \\
    \end{aligned}
\end{equation}
From the relation $e^{-i\hat{H}t/\hbar} \ket{\text{v}} \ket{n}_m = e^{-i n\mu t} \ket{\text{v}} \ket{n}_m$, we then calculate for any $n_\text{i,f}$~\cite{Caneva2015, Chang2016}
\begin{equation}
    \begin{aligned}
        &_{m}\bra{n_\text{f}}\bra{\text{v}}\hat{c}_\text{H}(t_1)\hat{c}_\text{H}^\dagger(t_2)\ket{\text{v}} \ket{n_\text{i}}_m \\
        &= \Tr\ab[e^{i\hat{H}t_1/\hbar} \hat{c} e^{-i\hat{H}(t_1-t_2)/\hbar} \hat{c}^\dagger e^{-i\hat{H}t_2/\hbar} \ketbra{\text{v}}{\text{v}} \otimes \ketbra{n_\text{i}}{n_\text{f}}_m] \\
        &= \Tr\ab[\hat{c} e^{-i\hat{H}(t_1-t_2)/\hbar} \hat{c}^\dagger \ketbra{\text{v}}{\text{v}} \otimes \ketbra{n_\text{i}}{n_\text{f}}_m e^{i\hat{H}t_1/\hbar}] e^{-i n_\text{i} \mu t_2} \\
        &= \Tr\ab[\hat{c} e^{-i\hat{H}(t_1-t_2)/\hbar} \hat{c}^\dagger \ketbra{\text{v}}{\text{v}} \otimes \ketbra{n_\text{i}}{n_\text{f}}_m e^{i\hat{H}(t_1-t_2)/\hbar}] e^{i (n_\text{f} -n_\text{i}) \mu t_2} \\
        &= \Tr_{\text{sys},m}\ab[\hat{c} e^{\mathcal{L}(t_1-t_2)} \hat{c}^\dagger \ketbra{g,0}{g,0} \otimes \ketbra{n_\text{i}}{n_\text{f}}_m ] e^{i (n_\text{f} -n_\text{i}) \mu t_2},
    \end{aligned}
\end{equation}
where $\ket{g,0} = \ket{g}_a\ket{0}_c$.
Here, $\mathcal{L}$ is the Liouville superoperator that satisfies
\begin{equation}
    \mathcal{L}[\cdot] = -\frac{i}{\hbar} \ab[\hat{\mathcal{H}}_\text{sys}+ \hat{H}_\text{ho}, \cdot] + 2\kappa_\text{ex} \hat{c} \cdot \hat{c}^\dagger + 2\kappa_\text{in} \hat{c} \cdot \hat{c}^\dagger + 2\gamma \hat{\sigma} \cdot \hat{\sigma}^\dagger,
\end{equation}
where $\hat{\mathcal{H}}_\text{sys}$ is the (non-Hermitian) effective system Hamiltonian,
\begin{equation}
    \hat{\mathcal{H}}_\text{sys} = \hat{H}_\text{sys} - i\hbar (\gamma \ketbra{e}{e}_a + \kappa \hat{c}^\dagger \hat{c}).
\end{equation}
Considering the initial condition, we find
\begin{equation}
    \begin{aligned}
        &e^{\mathcal{L}t} \hat{c}^\dagger \ketbra{g,0}{g,0} \otimes \ketbra{n_\text{i}}{n_\text{f}}_m  \\
        &= e^{-i (\hat{\mathcal{H}}_\text{sys}+ \hat{H}_\text{ho})t/\hbar} \hat{c}^\dagger \ketbra{g,0}{g,0} \otimes \ketbra{n_\text{i}}{n_\text{f}}_m e^{i (\hat{\mathcal{H}}_\text{sys}+ \hat{H}_\text{ho})t/\hbar}  \\
        &= e^{-i (\hat{\mathcal{H}}_\text{sys}+ \hat{H}_\text{ho})t/\hbar} \hat{c}^\dagger \ketbra{g,0}{g,0} \otimes \ketbra{n_\text{i}}{n_\text{f}}_m e^{in_\text{f}\mu t},
    \end{aligned}
\end{equation}
leading to
\begin{equation}
    \begin{aligned}
        &_{m}\bra{n_\text{f}}\bra{\text{v}}\hat{c}_\text{H}(t_1)\hat{c}_\text{H}^\dagger(t_2)\ket{\text{v}} \ket{n_\text{i}}_m \\
        &=\ _{m}\bra{n_\text{f}}\bra{\text{v}} e^{in_\text{f}\mu t_1} \hat{c} e^{-i (\hat{\mathcal{H}}_\text{sys}+ \hat{H}_\text{ho})(t_1-t_2)/\hbar} \hat{c}^\dagger e^{-i n_\text{i}\mu t_2}   \ket{\text{v}} \ket{n_\text{i}}_m.
    \end{aligned}
\end{equation}
This means
\begin{equation} \label{ap_eq:second_term_calc_2}
    \begin{aligned}
        &\bra{\text{v}}\hat{c}_\text{H}(t_1)\hat{c}_\text{H}^\dagger(t_2)\ket{\text{v}} \\
        &= e^{i\hat{n}_b \mu t_1} \bra{\text{v}} \hat{c} e^{-i (\hat{\mathcal{H}}_\text{sys}+ \hat{H}_\text{ho})(t_1-t_2)/\hbar} \hat{c}^\dagger \ket{\text{v}} e^{-i\hat{n}_b \mu t_2}.
    \end{aligned}
\end{equation}
Thus, by substituting Eqs.~\eqref{ap_eq:second_term_calc_1} and \eqref{ap_eq:second_term_calc_2} into Eq.~\eqref{ap_eq:S_Delta_f_Delta_i}, we find
\begin{equation}
    \begin{aligned}
        &\hat{\mathcal{S}}_{\Delta_\text{f},\Delta_i} \\
        &= \delta(\Delta_\text{f}-\Delta_\text{i}) \hat{I}_m  \\
        & \hspace{1em} -\frac{\kappa_\text{ex}}{\pi} \int_{-\infty}^{\infty} \dd{t_1} e^{i(\hat{n}_b \mu + \Delta_\text{f}) t_1} \\
        & \hspace{1.5em} \times \int_{-\infty}^{t_1} \dd{t_2}  \bra{\text{v}} \hat{c} e^{-i (\hat{\mathcal{H}}_\text{sys}+ \hat{H}_\text{ho})(t_1-t_2)/\hbar} \hat{c}^\dagger \ket{\text{v}} e^{-i(\hat{n}_b \mu + \Delta_\text{i})t_2}.
    \end{aligned}
\end{equation}

To further calculate the second term, we now consider the $S$ matrix
\begin{equation}
    \mathcal{S}_{\Delta_\text{f},\Delta_\text{i}}^{n_\text{f},n_\text{i}} =\ _{m}\bra{n_\text{f}}\hat{\mathcal{S}}_{\Delta_\text{f},\Delta_\text{i}}\ket{n_\text{i}}_{m}.
\end{equation}
From the relation
\begin{equation}
    \begin{aligned}
        &\int_{-\infty}^{t_1} \dd{t_2}  \bra{\text{v}} \hat{c} e^{-i (\hat{\mathcal{H}}_\text{sys}+ \hat{H}_\text{ho})(t_1-t_2)/\hbar} \hat{c}^\dagger \ket{\text{v}} e^{-i(n_\text{i} \mu + \Delta_\text{i})t_2} \\
        &= i\hbar \bra{\text{v}} \hat{c} \hat{G}(\hbar(n_\text{i} \mu + \Delta_\text{i})) \hat{c}^\dagger \ket{\text{v}} e^{-i(n_\text{i} \mu + \Delta_\text{i})t_1},
    \end{aligned}
\end{equation}
where
\begin{equation}
    \hat{G}(z) = \frac{1}{z- (\hat{\mathcal{H}}_\text{sys} + \hat{H}_\text{ho})},
\end{equation}
is the resolvent of $\hat{\mathcal{H}}_\text{sys} + \hat{H}_\text{ho}$~\cite{Atom-photon.ch3}, we obtain
\begin{equation} \label{ap_eq:S_matrix_1}
    \begin{aligned}
        &\mathcal{S}_{\Delta_\text{f},\Delta_\text{i}}^{n_\text{f},n_\text{i}} \\
        &= \delta(\Delta_\text{f}-\Delta_\text{i}) \delta_{n_\text{f},n_\text{i}}  \\
        & \hspace{1em} -\frac{i\hbar \kappa_\text{ex}}{\pi}\ _m\bra{n_\text{f}}\bra{\text{v}} \hat{c} \hat{G}(\hbar(n_\text{i} \mu + \Delta_\text{i})) \hat{c}^\dagger \ket{\text{v}}\ket{n_\text{i}}_m  \\
        & \hspace{2em} \times\int_{-\infty}^{\infty} \dd{t_1} e^{i [\Delta_\text{f}-\Delta_\text{i}+(n_\text{f}-n_\text{i})\mu] t_1} \\
        &=\delta(\Delta_\text{f}-\Delta_\text{i}) \delta_{n_\text{f},n_\text{i}} \\
        &\hspace{1em} - \delta(\Delta_\text{f}-\Delta_\text{i}+(n_\text{f}-n_\text{i})\mu) \\
        &\hspace{3em} \times 2i\hbar \kappa_\text{ex} \ _m\bra{n_\text{f}}\bra{\text{v}} \hat{c} \hat{G}(\hbar(n_\text{i} \mu + \Delta_\text{i})) \hat{c}^\dagger \ket{\text{v}}\ket{n_\text{i}}_m.
    \end{aligned}
\end{equation}

\paragraph{Calculation of the resolvent.} Here, we calculate $\bra{\text{v}} \hat{c} \hat{G}(z)  \hat{c}^\dagger\ket{\text{v}}$ to derive a simple formula of $\hat{\mathcal{S}}_{\Delta_\text{f},\Delta_\text{i}}$.
We first decompose the effective Hamiltonian as
\begin{equation}
    \hat{\mathcal{H}}_\text{sys} + \hat{H}_\text{ho} = \hat{\mathcal{H}}_0 + \hat{V},
\end{equation}
where
\begin{equation}
    \begin{aligned}
        \hat{\mathcal{H}}_0 =& \ketbra{g}{g}_a \otimes \hbar\mu \hat{n}_b  \\
        &+ \ketbra{e}{e}_a \otimes \hbar \ab(\mu_e \hat{S}_b^\dagger(\zeta) \hat{n}_b \hat{S}_b(\zeta)  + \Delta_a + \frac{\Delta \mu}{2} - i\gamma) \\
        &-i\hbar \kappa \hat{c}^\dagger \hat{c}.
    \end{aligned}
\end{equation}
Here, we have defined the squeezing operator in the motion subspace
\begin{equation}
    \hat{S}_b(r) = \exp\ab[\frac{r}{2}(\hat{b}^2 - \hat{b}^{\dagger 2})],
\end{equation}
which satisfies
\begin{equation}
    \hat{S}_b^\dagger(r) \mqty{\hat{q} \\ \hat{p}} \hat{S}_b(r) = \mqty{e^{-r}\hat{q} \\ e^{r}\hat{p}},
\end{equation}
and
\begin{equation} \label{ap_eq:def_zeta}
    \zeta = -\frac{1}{2} \ln\ab(\frac{\mu_e}{\mu}).
\end{equation}
From the relation
\begin{equation} \label{ap_eq:general_inverse_relation}
    \frac{1}{\hat{A}}-\frac{1}{\hat{B}} = \frac{1}{\hat{A}}(\hat{B}-\hat{A})\frac{1}{\hat{B}} = \frac{1}{\hat{B}}(\hat{B}-\hat{A})\frac{1}{\hat{A}},
\end{equation}
for any invertible operators $\hat{A}$ and $\hat{B}$, we find
\begin{equation} \label{eq:G_G_0}
    \hat{G}(z) = \hat{G}_0(z) + \hat{G}(z) \hat{V} \hat{G}_0(z) = \hat{G}_0(z) + \hat{G}_0(z) \hat{V} \hat{G}(z),
\end{equation}
where $\hat{G}_0(z) = 1/(z-\hat{\mathcal{H}}_0)$ is the resolvent of $\hat{\mathcal{H}}_0$.
From the relation
\begin{equation}
    \hat{G}_0(z) \hat{c}^\dagger \ket{\text{v}} = \hat{c}^\dagger \ket{\text{v}} \hat{\mathcal{C}}(z),
\end{equation}
where
\begin{equation}
    \hat{\mathcal{C}}(z) = \frac{1}{z - \hbar(\mu \hat{n}_b -i\kappa)},
\end{equation}
we then find
\begin{equation}
    \begin{aligned}
        \bra{\text{v}}\hat{c} \hat{G}(z) \hat{c}^\dagger\ket{\text{v}} =& \hat{\mathcal{C}}(z) + \bra{\text{v}}\hat{c} \hat{G}(z)\hat{V} \hat{c}^\dagger\ket{\text{v}}\hat{\mathcal{C}}(z) \\
        =& \hat{\mathcal{C}}(z) + \hat{\mathcal{C}}(z)\bra{\text{v}}\hat{c} \hat{T}(z) \hat{c}^\dagger\ket{\text{v}}\hat{\mathcal{C}}(z),
    \end{aligned}
\end{equation}
where
\begin{equation}
    \hat{T}(z) = \hat{V} + \hat{V}\hat{G}(z)\hat{V}.
\end{equation}
Furthermore, Eq.~\eqref{eq:G_G_0} gives the relation
\begin{equation}
    \hat{T}(z) = \hat{V} + \hat{V}\hat{G}_0(z) \hat{V} + \hat{V}\hat{G}_0(z) \hat{V}\hat{G}_0(z) \hat{V} + \cdots,
\end{equation}
leading to
\begin{equation}
    \begin{aligned}
        &\bra{\text{v}}\hat{c} \hat{T}(z) \hat{c}^\dagger\ket{\text{v}} \\
        &= \bra{\text{v}}\hat{c} [\hat{V}\hat{G}_0(z)\hat{V} + \hat{V}\hat{G}_0(z) \hat{V}\hat{G}_0(z) \hat{V}\hat{G}_0(z)\hat{V} + \cdots] \hat{c}^\dagger\ket{\text{v}},
    \end{aligned}
\end{equation}
where a term including an odd number of $\hat{V}$ is reduced to zero because $\hat{V}$ maps $\hat{c}\ket{\text{v}}$ to $\hat{\sigma}^\dagger \ket{\text{v}}$ and vice versa.
To simplify this, we first calculate
\begin{equation}
    \hat{G}_0(z) \hat{\sigma}^\dagger \ket{\text{v}} = \hat{\sigma}^\dagger \ket{\text{v}} \hat{S}_b^\dagger(\zeta) \hat{\mathcal{A}}(z) \hat{S}_b(\zeta),
\end{equation}
where
\begin{equation}
    \hat{\mathcal{A}}(z) = \frac{1}{z - \hbar\ab[\mu\hat{n}_b + \Delta \mu (\hat{n}_b + 1/2) +\Delta_a -i\gamma]}.
\end{equation}
This leads to
\begin{equation} \label{ap_eq:VG_0Vc^dag|v>}
    \begin{aligned}
        \hat{V}\hat{G}_0(z) \hat{V} \hat{c}^\dagger\ket{\text{v}} =&\hat{V}\hat{G}_0(z) \hat{\sigma}^\dagger\ket{\text{v}} \cdot \hbar g\hat{d} \\
        =& \hat{V} \hat{\sigma}^\dagger\ket{\text{v}} \cdot \hat{S}_b^\dagger(\zeta) \hat{\mathcal{A}}(z) \hat{S}_b(\zeta) \hbar g\hat{d} \\
        =& \hat{c}^\dagger\ket{\text{v}} \cdot (\hbar g)^2 \hat{d}^\dagger \hat{S}_b^\dagger(\zeta) \hat{\mathcal{A}}(z) \hat{S}_b(\zeta) \hat{d}.
    \end{aligned}
\end{equation}
This gives
\begin{equation}
    \begin{aligned}
        &\bra{\text{v}}\hat{c} \hat{T}(z) \hat{c}^\dagger\ket{\text{v}} \\
        &= (\hbar g)^2 \hat{d}'^\dagger \hat{\mathcal{A}}(z)\hat{d}' + (\hbar g)^4 \hat{d}'^\dagger \hat{\mathcal{A}}(z) \hat{d}' \hat{\mathcal{C}}(z) \hat{d}'^\dagger \hat{\mathcal{A}}(z) \hat{d}' + \cdots,
    \end{aligned}
\end{equation}
where $\hat{d}' = \hat{S}_b(\zeta) \hat{d}$, resulting in
\begin{equation} \label{ap_eq:resolvent_relation}
    \begin{aligned}
        \bra{\text{v}}\hat{c} \hat{G}(z) \hat{c}^\dagger\ket{\text{v}} =& \hat{\mathcal{C}}(z) + (\hbar g)^2\hat{\mathcal{C}}(z) \hat{d}'^\dagger \hat{\mathcal{A}}(z)\hat{d}' \hat{\mathcal{C}}(z) + \cdots \\
        =& \hat{\mathcal{R}}(z)\hat{\mathcal{C}}(z),
    \end{aligned}
\end{equation}
where
\begin{equation}
    \begin{aligned}
        \hat{\mathcal{R}}(z) =& \hat{I}_m + (\hbar g)^2\hat{\mathcal{C}}(z) \hat{d}'^\dagger \hat{\mathcal{A}}(z)\hat{d}' + \cdots \\
        =& \ab[\hat{I}_m - (\hbar g)^2\hat{\mathcal{C}}(z) \hat{d}^\dagger \hat{S}_b^\dagger(\zeta) \hat{\mathcal{A}}(z) \hat{S}_b(\zeta) \hat{d}]^{-1}.
    \end{aligned}
\end{equation}

For the $S$ matrix $\mathcal{S}_{\Delta_\text{f},\Delta_\text{i}}^{n_\text{f},n_\text{i}}$ in Eq.~\eqref{ap_eq:S_matrix_1}, we have
\begin{equation}
    \begin{aligned}
        &_m\bra{n_\text{f}}\bra{\text{v}} \hat{c} \hat{G}(\hbar(n_\text{i} \mu + \Delta_\text{i})) \hat{c}^\dagger \ket{\text{v}}\ket{n_\text{i}}_m \\
        &= \mathcal{C}(\hbar \Delta_\text{i})\ _m\bra{n_\text{f}}\hat{\mathcal{R}}(\hbar(n_\text{i} \mu + \Delta_\text{i})) \ket{n_\text{i}}_m,
    \end{aligned}
\end{equation}
where $\mathcal{C}(z)$ is defined in Eq.~\eqref{eq:definition_of_C_A}.
This leads to
\begin{equation}
    \mathcal{S}_{\Delta_\text{f},\Delta_\text{i}}^{n_\text{f},n_\text{i}} = \delta(\Delta_\text{f}-\Delta_\text{i}+(n_\text{f}-n_\text{i})\mu) \ _m\bra{n_\text{f}} \hat{\mathcal{S}}(\Delta_\text{i}) \ket{n_\text{i}}_m,
\end{equation}
where we have defined
\begin{equation}
    \hat{\mathcal{S}}(z) = \hat{I}_m - \frac{2\kappa_\text{ex}}{\kappa-i z} \sum_{n} \hat{\mathcal{R}}(\hbar z + \hbar n\mu) \ketbra{n}{n}_m.
\end{equation}

\paragraph{Compact operator-based form of motion-photon interaction.}
In the following, we derive a compact operator-based form of motion-photon interaction, by focusing on the single-photon subspace of the propagating mode.
In the subspace, the unitary propagator $\hat{U}_\text{I}$ [Eq.~\eqref{eq:propagator}, with $t_\text{f}\to \infty, t_\text{i}\to -\infty$] reduces to
\begin{equation}
    \begin{aligned}
        & \iint \odif{\Delta_\text{f}}\dd{\Delta_\text{i}} \hat{a}^\dagger(\Delta_\text{f}) \ket{\text{v}}\hat{\mathcal{S}}_{\Delta_\text{f},\Delta_\text{i}} \bra{\text{v}} \hat{a}(\Delta_\text{i})  \\
        &= \sum_{n_\text{f},n_\text{i}} \int \dd{\Delta} \hat{a}^\dagger(\Delta-(n_\text{f}-n_\text{i})\mu)\ketbra{\text{v}}{\text{v}} \hat{a}(\Delta) \\
        &\hspace{3em} \times \ketbra{n_\text{f}}{n_\text{f}}_m \hat{\mathcal{S}}( \Delta) \ketbra{n_\text{i}}{n_\text{i}}_m.
    \end{aligned}
\end{equation}
By introducing a frequency shift operator in up to one photon subspace~\cite{Fabre2020, Fabre2022}
\begin{equation}
    \hat{\mathcal{D}}(\omega) = \int \dd{\omega^\prime} \hat{a}^\dagger(\omega^\prime + \omega) \hat{a}(\omega^\prime),
\end{equation}
which satisfies $\hat{\mathcal{D}}(\omega) \hat{a}^\dagger(\Delta)\ket{\emptyset}_p = \hat{a}^\dagger(\Delta + \omega)\ket{\emptyset}_p$, we rewrite the above expression as
\begin{equation}
    \begin{aligned}
        &\iint \odif{\Delta_\text{f}}\dd{\Delta_\text{i}} \hat{a}^\dagger(\Delta_\text{f}) \ket{\text{v}}\hat{\mathcal{S}}_{\Delta_\text{f},\Delta_\text{i}} \bra{\text{v}} \hat{a}(\Delta_\text{i}) \\
        &= \int \dd{\Delta} \hat{\mathcal{K}}(\Delta) \hat{a}^\dagger(\Delta) \hat{a}(\Delta),
    \end{aligned}
\end{equation}
in the single-photon subspace, where
\begin{equation}
    \hat{\mathcal{K}}(\Delta) = \sum_{n_\text{f},n_\text{i}}\ketbra{n_\text{f}}{n_\text{f}}_m \hat{\mathcal{D}}(-(n_\text{f}-n_\text{i})\mu) \hat{\mathcal{S}}( \Delta) \ketbra{n_\text{i}}{n_\text{i}}_m.
\end{equation}
From the relation
\begin{equation} \label{relation_of_D(omega)}
    \begin{aligned}
        &\hat{\mathcal{D}}(\omega)\hat{\mathcal{D}}(\omega^\prime) \\
        &= \hat{\mathcal{D}}(\omega+\omega^\prime)\\
        &\hspace{1em}+ \iint \odif{\Delta}\dd{\Delta^\prime} \hat{a}^\dagger(\Delta+\omega) \hat{a}^\dagger(\Delta^\prime + \omega^\prime) \hat{a}(\Delta) \hat{a}(\Delta^\prime),
    \end{aligned}
\end{equation}
we find that $\hat{\mathcal{D}}(\omega_1)\hat{\mathcal{D}}(\omega_2) = \hat{\mathcal{D}}(\omega_1+\omega_2)$ in the subspace.
By using this, we find
\begin{equation}
    \begin{aligned}
        \hat{\mathcal{K}}(\Delta) =& \sum_{n_\text{f},n_\text{i}} \hat{\mathcal{D}}(-n_\text{f}\mu)\ketbra{n_\text{f}}{n_\text{f}}_m \hat{\mathcal{S}}(\Delta) \ketbra{n_\text{i}}{n_\text{i}}_m \hat{\mathcal{D}}(n_\text{i}\mu) \\
        =& \hat{\mathcal{D}}(-\mu \hat{n}_b)\hat{\mathcal{S}}(\Delta) \hat{\mathcal{D}}(\mu \hat{n}_b),
    \end{aligned}
\end{equation}
where we have defined $\hat{O}(\hat{n}_b) = \sum_{n} \hat{O}(n)\ketbra{n}[_m]{n}$ for any operator $\hat{O}(z)$.

\paragraph{Three-dimensional motion space.}
Our results can be straightforwardly extended to a three-dimensional motion space.
In the following, we set $\hbar = 1$ unless otherwise specified.
For the three-dimensional case, the harmonic-oscillator potential term is given by
\begin{equation}
    \begin{aligned}
        \hat{H}_\text{ho} = \sum_{\alpha \in \{x,y,z\}}& \ketbra{g}{g}_a \otimes \mu^\alpha (\hat{b}^\alpha)^\dagger\hat{b}^\alpha \\
        &+ \ketbra{e}{e}_a \otimes \ab[\mu_e^\alpha (\hat{b}_e^\alpha)^\dagger\hat{b}^\alpha_e + \frac{\mu^\alpha_e-\mu^\alpha}{2}],
    \end{aligned}
\end{equation}
where the superscript $\alpha \in \{x,y,z\}$ represents the axis to which the variable corresponds.
By replacing the operators $\mu \hat{n}_b \to \sum_{\alpha} \mu^\alpha \hat{n}_b^\alpha$, $\mu_e \hat{n}_b \to \sum_{\alpha} \mu_e^\alpha \hat{n}_b^\alpha$, and $\hat{S}_b(\zeta) \to \prod_{\alpha} \hat{S}_b^\alpha(\zeta^\alpha)$, the discussion in the one-dimensional case still holds, leading to the $S$ matrix as
\begin{equation}
    \begin{aligned}
        &\mathcal{S}_{\Delta_\text{f},\Delta_\text{i}}^{\{n_\text{f}^\alpha\},\{n_\text{i}^\alpha\}} \\
        &= \delta\ab(\Delta_\text{f}-\Delta_\text{i}+  \sum_{\alpha} (n_\text{f}^\alpha-n_\text{i}^\alpha)\mu^\alpha) \ _m\bra{\{n_\text{f}^\alpha\}} \hat{\mathcal{S}}(\Delta) \ket{\{n_\text{i}^\alpha\}}_m,
    \end{aligned}
\end{equation}
where
\begin{equation}
    \hat{\mathcal{S}}(\Delta) =  \hat{I}^{\otimes 3}_m - \frac{2\kappa_\text{ex}}{\kappa-i\Delta}\sum_{\{n^\alpha\}} \hat{R}\ab(\Delta + \sum_{\alpha}\mu^\alpha n^{\alpha}) \ketbra{\{n_\text{f}^\alpha\}}{\{n_\text{i}^\alpha\}}_m.
\end{equation}

\supsubsection{Case of multiple two-level systems} \label{ap_subsec_N_two-level_system_case}

Here, we consider $N$ two-level systems coupled to a single cavity, where parameters and operators for the $j$th two-level system $(j\in\{1,2,\cdots,N\})$ are labeled by the subscript $j$.
For simplicity, we again consider a one-dimensional motion space, while a three-dimensional case can be handled in the same manner as the $N=1$ case.

The system and harmonic potential Hamiltonian are given by
\begin{equation}
    \hat{H}_\text{sys}^{(N)} = \sum_{j\in \mathsf{Q}} \hat{H}_{\text{sys},j}, \quad  \hat{H}_\text{ho}^{(N)} = \sum_{j\in \mathsf{Q}} \hat{H}_{\text{ho},j},
\end{equation}
where $\mathsf{Q} = \{1,2,\cdots,N\}$.
This gives the effective system Hamiltonian as
\begin{equation}
    \hat{\mathcal{H}}_\text{sys}^{(N)} = \sum_{j \in \mathsf{Q}} \ab(\hat{H}_{\text{sys},j} - i \gamma_j \ketbra{e}{e}_{a,j}) - i\kappa \hat{c}^\dagger\hat{c}.
\end{equation}
Here, we start with the $S$ matrix as
\begin{equation}
    \mathcal{S}_{\Delta_\text{f},\Delta_\text{i}}^{\bm{n}_\text{f}, \bm{n}_\text{i}} = \,_m\bra{\bm{n}_\text{f}}\hat{\mathcal{S}}_{\Delta_\text{f},\Delta_\text{i}}\ket{\bm{n}_\text{i}}_m,
\end{equation}
where $\bm{n} = (n_1,n_2,\cdots,n_N)$ and $\ket{\bm{n}}_m = \ket{n_1}_{m,1}\ket{n_2}_{m,2}\cdots\ket{n_N}_{m,N}$.
Following a similar procedure as in the $N=1$ case in Sec.~\ref{ap_subsec:sc_theory_single_two-level_system}, the $S$ matrix reads
\begin{equation} \label{ap_eq:S_matrix_N}
    \begin{aligned}
        &\mathcal{S}_{\Delta_\text{f},\Delta_\text{i}}^{\bm{n}_\text{f}, \bm{n}_\text{i}} \\
        &=\delta(\Delta_\text{f}-\Delta_\text{i}) \delta_{\bm{n}_\text{f},\bm{n}_\text{i}} \\
        &\hspace{1em} - \delta(\Delta_\text{f}-\Delta_\text{i}+\bm{\mu}^\top(\bm{n}_\text{f}-\bm{n}_\text{i})) \\
        &\hspace{3em} \times 2i\hbar \kappa_\text{ex} \ _m\bra{\bm{n}_\text{f}} \bra{\text{v}} \hat{c} \hat{G}^{(N)}(\Delta + \bm{\mu}^\top \bm{n}_\text{i}) \hat{c}^\dagger \ket{\text{v}} \ket{\bm{n}_\text{i}}_m,
    \end{aligned}
\end{equation}
where $\hat{G}^{(N)} = 1/[z-(\hat{\mathcal{H}}_\text{sys}^{(N)} + \hat{H}_\text{ho}^{(N)})]$.

For the calculation of the resolvent $\bra{\text{v}}\hat{c}\hat{G}^{(N)}(z)\hat{c}^\dagger \ket{\text{v}}$, the main difference from the $N=1$ case appears in the term corresponding to Eq.~\eqref{ap_eq:VG_0Vc^dag|v>};
by introducing
\begin{equation}
    \hat{G}_0^{(N)}(z) = \frac{1}{z-\ab[\hat{\mathcal{H}}_\text{sys}^{(N)} + \hat{H}_\text{ho}^{(N)} -\hat{V}^{(N)}]},
\end{equation}
where
\begin{equation}
    \hat{V}^{(N)} = \sum_{j\in \mathsf{Q}} g_j(\hat{c}\hat{\sigma}_j^\dagger \hat{d}_j + \text{h.c.}),
\end{equation}
we find
\begin{equation}
    \begin{aligned}
        &\hat{V}^{(N)}\hat{G}_0^{(N)}(z)\hat{V}^{(N)}\hat{c}^\dagger \ket{\text{v}} \\
        &=  \sum_{j\in \mathsf{Q}} \hat{V}^{(N)}\hat{G}_0^{(N)}(z) \hat{\sigma}_j^\dagger\ket{\text{v}} \cdot g_j \hat{d}_j \\
        &= \sum_{j\in \mathsf{Q}} \hat{V}^{(N)} \hat{\sigma}_j^\dagger\ket{\text{v}} \cdot \hat{S}_{b,j}^\dagger(\zeta_j) \hat{\mathcal{A}}_j^{(N)}(z) \hat{S}_{b,j}(\zeta_j) g_j \hat{d}_j \\
        &= \hat{c}^\dagger \ket{\text{v}} \sum_{j\in \mathsf{Q}} g_j^2 \hat{d}_j^\dagger \hat{S}_{b,j}^\dagger(\zeta_j) \hat{\mathcal{A}}_j^{(N)}(z) \hat{S}_{b,j}(\zeta_j)\hat{d}_j,
    \end{aligned}
\end{equation}
where we have defined
\begin{equation}
    \hat{\mathcal{A}}_j^{(N)}(z) = \ab\{z- \ab[\hat{H}_\text{ho,g}^{(N)} + \Delta \mu_j\ab(\hat{n}_{b,j} + \frac{1}{2}) + \Delta_{a,j} - i\gamma_j]\}^{-1},
\end{equation}
and
\begin{equation}
    \hat{H}_\text{ho,g}^{(N)} = \sum_{k\in \mathsf{Q}} \mu_k \hat{n}_{b,k}.
\end{equation}
Following the similar procedure as in the $N=1$ case, we then obtain
\begin{equation}
    \bra{\text{v}}\hat{c} \hat{G}^{(N)}(z) \hat{c}^\dagger\ket{\text{v}} = \hat{\mathcal{R}}^{(N)}(z) \hat{\mathcal{C}}^{(N)}(z),
\end{equation}
where
\begin{equation}
    \begin{aligned}
        \hat{\mathcal{C}}^{(N)}(z) =& \ab[z- \ab(\hat{H}_\text{ho,g}^{(N)} -i\kappa)]^{-1}, \\
    \end{aligned}
\end{equation}
and
\begin{equation}
    \begin{aligned}
        &\hat{\mathcal{R}}^{(N)}(z) \\
        &= \ab[\hat{I}_m^{(N)}-\hat{\mathcal{C}}^{(N)}(z) \sum_{j\in \mathsf{Q}} g_j^2 \hat{d}_j^\dagger \hat{S}_{b,j}^\dagger(\zeta_j) \hat{\mathcal{A}}_j^{(N)}(z)\hat{S}_{b,j}(\zeta_j) \hat{d}_j]^{-1}.
    \end{aligned}
\end{equation}
For the $S$ matrix $\mathcal{S}_{\Delta_\text{f},\Delta_\text{i}}^{\bm{n}_\text{f}, \bm{n}_\text{i}}$ in Eq.~\eqref{ap_eq:S_matrix_N}, we have
\begin{equation}
    \begin{aligned}
        &_m\bra{\bm{n}_\text{f}} \bra{\text{v}} \hat{c} \hat{G}^{(N)}(\Delta + \bm{\mu}^\top \bm{n}_\text{i}) \hat{c}^\dagger \ket{\text{v}} \ket{\bm{n}_\text{i}}_m \\
        &=  \mathcal{C}(\Delta)\ _m\bra{\bm{n}_\text{f}} \hat{\mathcal{R}}^{(N)}(\Delta + \bm{\mu}^\top \bm{n}_\text{i}) \ket{\bm{n}_\text{i}}_m.
    \end{aligned}
\end{equation}
This leads to
\begin{equation}
    \mathcal{S}_{\Delta_\text{f},\Delta_\text{i}}^{\bm{n}_\text{f}, \bm{n}_\text{i}} =\delta(\Delta_\text{f}-\Delta_\text{i}+\bm{\mu}^\top(\bm{n}_\text{f}-\bm{n}_\text{i})) \ _m\bra{\bm{n}_\text{f}} \hat{\mathcal{S}}^{(N)}(\Delta) \ket{\bm{n}_\text{i}}_m,
\end{equation}
where
\begin{equation}
    \hat{\mathcal{S}}^{(N)}(\Delta) = \hat{I}_m^{(N)} - \frac{2\kappa_\text{ex}}{\kappa-i\Delta} \sum_{\bm{n}} \hat{\mathcal{R}}^{(N)}(\Delta + \bm{\mu}^\top \bm{n}) \ketbra{\bm{n}}{\bm{n}}_m.
\end{equation}
We finally note that the compact operator-based form of motion-photon interaction can be obtained as needed, by following a similar procedure as in the $N=1$ case.

\supsection{Dispersive coupling regime in a standing-wave cavity} \label{ap_sec:dispersive_coupling}

For the regime of dispersive coupling between a two-level system and a standing-wave cavity, where the input photon is largely detuned from both the two-level system and the cavity, Ref.~\cite{Neumeier2018_PRA, Neumeier2018_IOP} presented effective cavity detuning and linewidth by assuming adiabatic elimination of the excited state.
In contrast, our scattering theory does not assume adiabatic elimination, which makes it applicable over a wide parameter range, but still reproduces the same detuning and linewidth by applying the equivalent approximation, as explained in the following.

Here, we consider $\mu_e = \mu$ for simplicity.
For the operator
\begin{equation}
    \hat{\mathcal{A}}(\Delta + n \mu ) = \frac{1}{\Delta-\Delta_a + i\gamma + \mu (n-\hat{n}_b)},
\end{equation}
a sufficiently large detuning such that $|\Delta-\Delta_a + i\gamma| \gg \mu \aab{n-\hat{n}_b}$ allows us to approximate $\hat{\mathcal{A}}(\Delta + n \mu ) \approx 1/(\Delta-\Delta_a + i\gamma)$, which means that we neglect the free oscillation of the motional state at frequency $\mu$ while the atom is transiently excited.
Similarly, we further assume $|\Delta + i\kappa|\gg \mu \aab{n-\hat{n}_b}$, resulting in $\hat{\mathcal{C}}(\Delta + n \mu) \approx 1/(\Delta+i\kappa)$.
This leads to
\begin{equation}
    \hat{\mathcal{S}}(\Delta) \approx \hat{I}_m - 2i\kappa_\text{ex}\ab(\Delta+i\kappa - \frac{g^2 \hat{d}^\dagger \hat{d}}{\Delta-\Delta_a+i\gamma})^{-1}.
\end{equation}
For the standing-wave cavity, $\hat{d}_\text{st} = u(\hat{q})$, where $u(\hat{q})$ is a Hermitian operator depending on the position $\hat{q}$ of the two-level system, we find
\begin{equation}
    \hat{\mathcal{S}}(\Delta) \approx \hat{I}_m - \frac{2i\kappa_\text{ex}}{\Delta - \delta \omega_c(\hat{q}) + i[\kappa + \delta \kappa(\hat{q})]},
\end{equation}
where the resonant frequency of the cavity is effectively shifted by
\begin{equation}
    \delta \omega_c(\hat{q}) = \frac{g^2(\Delta-\Delta_a)u^2(\hat{q})}{(\Delta-\Delta_a)^2 + \gamma^2},
\end{equation}
and the cavity linewidth is effectively broadened by
\begin{equation}
    \delta\kappa(\hat{q}) = \gamma \frac{g^2 u^2(\hat{q})}{(\Delta-\Delta_a)^2 + \gamma^2}.
\end{equation}
These results are the same as those in Refs.~\cite{Neumeier2018_PRA, Neumeier2018_IOP}.

\supsection{Atom-photon CZ gates} \label{ap_sec:phase_inversion}

For a polarization-encoded photonic qubit, photon reflection off a cavity hosting an atomic qubit, mediated by the linear optics shown in Fig.~\ref{fig2}(a), yields an atom-photon controlled-phase (CZ) gate.
The detailed mechanism can be found in many previous studies, such as Ref.~\cite{Duan2004, Raymer2024, Kikura2025_caps}.

Here, we focus on the atomic-state-dependent $\pi$ phase shift applied by photon reflection, as a fundamental block of the CZ gate.
For simplicity, we assume that the input photon is resonant with the cavity and contained within a sufficiently long pulse, allowing us to model a single-photon state as $\hat{a}^\dagger(\Delta =0)\ket{\emptyset}_p = \ket{\Delta = 0}_p$.
Before the detailed evaluation of the motion-photon coupling effect, we explain the target operation where we ignore the motion-photon coupling.
Here, the state $\ket{1}_a$ is resonantly coupled to the cavity, as shown in Fig.~\ref{fig2}(a).
For the initial atomic state in $\alpha\ket{0}_a + \beta \ket{1}_a \, (\vab{\alpha}^2 + \vab{\beta}^2 = 1)$, the atom-photon state after the photon reflection is given by
\begin{equation}
    [\alpha r_0(0) \ket{0}_a + \beta r_1(0) \ket{1}_a] \otimes \ket{\Delta=0}_p,
\end{equation}
where $r_i(\Delta)$ are the state-dependent reflection amplitudes,
\begin{equation}
    \begin{aligned}
        r_1(\Delta) =& r(\Delta) = 1- \frac{2\kappa_\text{ex}[\gamma-i(\Delta-\Delta_a)]}{(\kappa-i\Delta)[\gamma-i(\Delta-\Delta_a)] + g^2}, \\
        r_0(\Delta) =& r(\Delta)\Big|_{g=0} =1 -\frac{2\kappa_\text{ex}}{\kappa-i\Delta}.
    \end{aligned}
\end{equation}
The photon detection, which succeeds with a probability $p = \vab{\alpha r_0(0)  + \beta r_1(0)}^2$, projects the atom onto $(\alpha r_0(0) \ket{0}_a + \beta r_1(0) \ket{1}_a)/\sqrt{p}$.
Optimized output coupling rate reads $\kappa_\text{ex}^\text{opt} = \kappa_\text{in}\sqrt{1+2C_\text{in}}$, ensuring
\begin{equation}
    -r_0(0) = r_1(0) = 1- \frac{2}{1+\sqrt{1+2C_\text{in}}} \eqqcolon r^{\text{opt}},
\end{equation}
where $C_\text{in} = g^2/(2\kappa_\text{in}\gamma)$ is the internal cooperativity; then the photon detection probability is given by $p = (r^{\text{opt}})^2$, and the projected state reduces to $-\alpha \ket{0}_a + \beta \ket{1}_a$.
This realizes the state-dependent phase shift.

We now incorporate the motional DOF.
The initial composite state is given by
\begin{equation}
    \hat{\rho}_\text{ini} = \sum_{i,j \in \{0,1\}} C_{ij}\ketbra{i}[_a]{j} \otimes \hat{\rho}_m \otimes \ketbra{\Delta=0}[_p]{\Delta=0},
\end{equation}
where $\hat{\rho}_m$ represents the initial motional state, and coefficients $C_{ij}$ of the initial atomic state satisfy $\Tr[\hat{\rho}_\text{ini}] = C_{00} + C_{11} = 1$.
The operator that maps the initial state to the reflected one is
\begin{equation}
    \hat{E} = \sum_{i = 0,1} \ketbra{i}{i}_a \otimes \int \dd{\Delta} \hat{\mathcal{K}}_i(\Delta) \hat{a}^\dagger(\Delta) \hat{a}(\Delta),
\end{equation}
where we have defined
\begin{gather}
    \hat{\mathcal{K}}_i(\Delta) = \hat{\mathcal{D}}(-\mu \hat{n}_b) \hat{\mathcal{S}}_{i}(\Delta) \hat{\mathcal{D}}(\mu \hat{n}_b) \quad (i\in \{0,1\}), \\
    \hat{\mathcal{S}}_1(\Delta) = \hat{\mathcal{S}}(\Delta), \quad \hat{\mathcal{S}}_0(\Delta) = \hat{\mathcal{S}}(\Delta)\Big|_{g=0} = r_0(\Delta) \hat{I}_m,
\end{gather}
for notation simplicity.
The unnormalized state after the photon reflection is then given by
\begin{equation}
    \begin{aligned}
        &\hat{E}\hat{\rho}_\text{ini}\hat{E}^\dagger \\
        &= \sum_{ij} C_{ij}\ketbra{i}[_a]{j} \otimes \hat{\mathcal{K}}_i(0) (\hat{\rho}_m \otimes \ketbra{\Delta=0}[_p]{\Delta=0}) \hat{\mathcal{K}}_j^\dagger(0).
    \end{aligned}
\end{equation}
Thus, the atom-motion state conditioned on the photon detection is given by
\begin{equation}
    \hat{\rho}_{am} = \frac{\Tr_p[\hat{E}\hat{\rho}_\text{ini}\hat{E}^\dagger]}{\Tr[\hat{E}\hat{\rho}_\text{ini}\hat{E}^\dagger]},
\end{equation}
where the denominator represents the photon detection probability,
\begin{equation}
    \begin{aligned}
        p =& \Tr[\hat{E}\hat{\rho}_\text{ini}\hat{E}^\dagger] \\
        =& \sum_{i\in\{0,1\}}C_{ii} \Tr_m[_p\bra{\Delta=0} \hat{\mathcal{K}}_i^\dagger(0)\hat{\mathcal{K}}_i(0)\ket{\Delta=0}_p \hat{\rho}_m].
    \end{aligned}
\end{equation}
From the relation in Eq.~\eqref{relation_of_D(omega)}, we find
\begin{equation}
    \begin{aligned}
        &_p\bra{\Delta=0} \hat{\mathcal{K}}_j^\dagger(0)\hat{\mathcal{K}}_i(0)\ket{\Delta=0}_p \\
        &={} _p\bra{\Delta=0}\hat{\mathcal{D}}(-\mu \hat{n}_b) \hat{\mathcal{S}}_{j}^\dagger(0) \hat{\mathcal{S}}_{i}(0) \hat{\mathcal{D}}(\mu \hat{n}_b)\ket{\Delta=0}_p \\
        &= \sum_n \ketbra{n}[_m]{n}\hat{\mathcal{S}}_{j}^\dagger(0) \hat{\mathcal{S}}_{i}(0)\ketbra{n}[_m]{n}.
    \end{aligned}
\end{equation}
This leads to
\begin{equation}
    p = \sum_{i=0,1}C_{ii} J_{ii},
\end{equation}
where
\begin{equation}
    J_{ij} = \sum_n {}_m\bra{n}\hat{\mathcal{S}}_{j}^\dagger(0) \hat{\mathcal{S}}_{i}(0)\ket{n}_m\braket*[3]{n}{\hat{\rho}_m}{n}_m.
\end{equation}
By tracing out the motional state, we obtain the projected atomic state as
\begin{equation}
    \hat{\rho}_a = \Tr_m[\hat{\rho}_{am}] = \frac{1}{p}\sum_{ij} C_{ij}J_{ij}\ketbra{i}{j}_a,
\end{equation}
which gives the fidelity of the phase shift.

\begin{figure}
    \centering
    \includegraphics[width=0.5\linewidth]{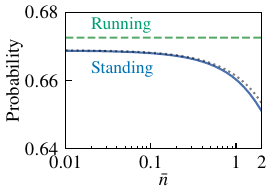}
    \caption{Success probability of a heralded state-dependent phase shift, with the same parameters as in Fig.~\ref{fig2}(b).
    The dotted line represents $(r^\text{opt})^2$ while replacing $C_\text{in}$ with $[1-(\bar{n}+1/2)\eta^2]^2 C_\text{in}$.}
    \label{fig4}
\end{figure}
While the main text provides the infidelity as a function of the mean phonon number $\bar{n}$ of the initial thermal state of atomic motion,
\begin{equation}
    \hat{\rho}_\text{th} = \sum_{n=0}^\infty \frac{\bar{n}^n}{(\bar{n}+1)^{n+1}} \ketbra{n}{n}_m,
\end{equation}
Fig.~\ref{fig4} shows the success probability of the state-dependent phase shift.
Although the probability for the running-wave cavity is nearly constant, the corresponding probability for the standing-wave cavity is slightly affected as $\bar{n}$ increases.
The latter is well captured by the analytical prediction $(r^\text{opt})^2$ with the damped coupling strength as
\begin{equation}
    g_{\bar{n}} =\ab[1-\ab(\bar{n}+\frac{1}{2})\eta^2]g,
\end{equation}
and can be explained by the perturbative expansion of the recoil operator for the standing-wave cavity.
Considering the case $\mu_e = \mu$ for simplicity, we expand the coupling operator up to the third order of $\eta$,
\begin{equation}
    \hat{d}_\text{st} = \hat{I}_m - \eta^2\ab(\hat{n}_b + \frac{1}{2}) - \eta^2 \frac{\hat{b}^2 + \hat{b}^{\dagger^2}}{2} + \mathcal{O}(\eta^4).
\end{equation}
This leads to
\begin{equation}
    \begin{aligned}
        \hat{d}_\text{st}\hat{\mathcal{A}}(z)\hat{d}_\text{st}^\dagger =& [\hat{I}_m - \eta^2(2\hat{n}_b + 1)]\hat{\mathcal{A}}(z) \\
        &-\frac{\eta^2}{2} \{\hat{b}^2 + \hat{b}^{\dagger^2}, \hat{\mathcal{A}}(z)\} + \mathcal{O}(\eta^4),
    \end{aligned}
\end{equation}
where $\{\hat{A}, \hat{B}\} = \hat{A}\hat{B} + \hat{B}\hat{A}$ is the anticommutator.
By using Eq.~\eqref{ap_eq:general_inverse_relation}, we find
\begin{equation}
    \begin{aligned}
        &\hat{R}_\text{st}(z) \\
        &= \hat{R}_{\text{st},0}(z) - \frac{g^2\eta^2}{2}\hat{R}_{\text{st},0}(z)\hat{\mathcal{C}}(z)\{\hat{b}^2 + \hat{b}^{\dagger^2}, \hat{\mathcal{A}}(z)\}\hat{R}_{\text{st},0}(z) \\
        &\hspace{0.5cm} + \mathcal{O}(\eta^4),
    \end{aligned}
\end{equation}
where
\begin{equation}
    \hat{R}_{\text{st},0}(z) = \ab\{\hat{I}_m - g^2[\hat{I}_m - \eta^2(2\hat{n}_b + 1)]\hat{\mathcal{C}}(z)\hat{\mathcal{A}}(z)\}^{-1}.
\end{equation}
This indicates that in the carrier transition, characterized by $\hat{R}_{\text{st},0}(z)$, the coupling strength is replaced with the effective coupling-strength operator,
\begin{equation}
    \hat{g}_\text{st} = \ab[\hat{I}_m - \ab(\hat{n}_b + \frac{1}{2})\eta^2]g,
\end{equation}
which is similar to the photon-emission case~\cite{Kikura2025_recoil}.

\supsubsection{Cavity-parameter dependence}
\label{ap_sec:cavity_parameter_dependence}

\begin{figure}
    \centering
    \includegraphics[width=\linewidth]{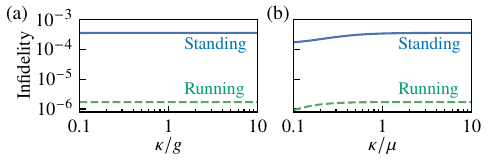}
    \caption{Infidelity of the phase gate obtained by scanning the parameters of atom-cavity systems.
    Here, we fix the internal cooperativity as $C_\text{in}=50.7$ [same as in Fig.~\ref{fig2}(b)] and adjust the outcoupling rate to satisfy $\kappa_\text{ex}^\text{opt} = \kappa_\text{in}\sqrt{1+2C_\text{in}}$.
    (a) Infidelity as a function of the ratio $\kappa/g$.
    (b) Infidelity with scanning the ratio $\kappa/\mu$.}
    \label{fig5}
\end{figure}

Here, we investigate the system-parameter dependence of the motion-induced infidelity of the phase shift, to compare it with the photon-emission process addressed in Ref.~\cite{Kikura2025_recoil}.
We first evaluate the infidelity by scanning the ratio $\kappa/g$, which classifies the atom-cavity system into two regimes, strong-coupling $(\kappa < g)$ and bad-cavity $(\kappa > g)$ regimes.
Figure~\ref{fig5}(a) shows that both cavities exhibit nearly constant infidelity, while the emission process suppresses the motion-photon interaction as $\kappa/g$ increases~\cite{Kikura2025_recoil}.
In Ref.~\cite{Kikura2025_recoil}, the strong-coupling regime $(\kappa/g < 1)$ allows the cavity photon to interact with the spin-motion system multiple times, leading to higher infidelity in remote entanglement generation.
In contrast, a sufficiently long-pulse photon in CAPS hardly excites the atom irrespective of the ratio $\kappa/g$, suppressing the motion-photon interaction.

We further investigate how the cavity's resolution of motional-sideband transitions influences the infidelity.
In Fig.~\ref{fig5}(b), we observe similar behavior to the photon-emission process~\cite{Kikura2025_recoil}: the resolved-sideband regime $(\kappa/\mu < 1)$ suppresses motion-induced infidelity.
In contrast, in the unresolved-sideband regime $(\kappa/\mu > 1)$, the cavity broadens the linewidth of the state $\ket{g}_a\ket{1}_c$ sufficiently that sideband transitions are allowed, even when the incident photon is resonant with the carrier transition.
We finally note that Ref.~\cite{Kikura2025_recoil} evaluates the motion-induced infidelity of remote entanglement generation involving two atoms, while we instead focus on the infidelity for a single atom-cavity system here;
the results in this work can, in general, be compared with those of Ref.~\cite{Kikura2025_recoil} by multiplying the reported infidelity by a factor of two. This is because we focus on the low-infidelity regime, where the total infidelity of the two-atom system is well approximated by the sum of the infidelities associated with the individual atoms.

\supsubsection{Analytical model for the infidelity arising from the state-dependent trapping potential}

\begin{figure}
    \centering
    \includegraphics[width=0.75\linewidth]{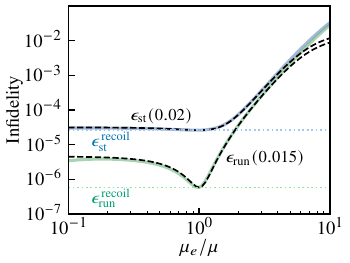}
    \caption{Error modeling for the state-dependent trapping potential, where we reproduce the infidelity of the running- and standing-wave cavities in Fig.~\ref{fig3}(b) with a wider range of $\mu_e/\mu$ (solid lines).
    The dotted lines represent the infidelities induced only by the recoil-kick effect, $\epsilon_\text{run}^\text{recoil}$ and $\epsilon_\text{st}^\text{recoil}$.
    The dashed lines represent the infidelity $\epsilon_\text{run(st)}(\Lambda)$ with the free parameter $\Lambda$ set to 0.015 and 0.02 for the running- and standing-wave cavities, respectively.}
    \label{fig6}
\end{figure}
We have investigated the effect arising from the state-dependent potential, as shown in Fig.~\ref{fig3}.
Here, we develop a simple analytical model to reproduce the numerical results in Fig.~\ref{fig3}(b).

First, we focus on the effect arising from the sequential motional effect consisting of squeezing, phase rotation, and antisqueezing, where the recoil-kick effect is ignored.
The resolvent $\hat{\mathcal{A}}(z)$ can be rewritten as
\begin{equation} \label{ap_eq:A_integral_form}
    \hat{\mathcal{A}}(z) = -i\int_0^\infty \dd{t} e^{-\gamma t} e^{-i[\mu_e \hat{n}_b + \Delta \mu/2 - (z-\Delta_a)]t},
\end{equation}
which includes the integration of the phase rotation.
Then, for the initial motional state to be the vacuum state, one cycle of the cyclic transition $\ket{g}_a\ket{1}_c \leftrightarrow \ket{e}\ket{0}_c$ changes the motional state to
\begin{equation}
    \hat{S}_b^\dagger(\zeta) \hat{\mathcal{A}}(\Delta) \hat{S}_b(\zeta) \ket{0}_m,
\end{equation}
where we ignore the recoil kick, $\hat{d}_\text{run}$ or $\hat{d}_\text{st}$.
Here, to capture the essence of the motional-state effect, we evaluate $|_m\bra{0}\hat{S}^\dagger(\zeta) e^{-i\mu_e \hat{n}_b t} \hat{S}(\zeta) \ket{0}_m|^2$, which is the overlap between the vacuum state and the motional state after the transition.
From the relation for the squeezed vacuum state,
\begin{equation}
    \hat{S}_b(\zeta) \ket{0}_m = \frac{1}{\sqrt{\cosh \zeta}} \sum_{n=0}^{\infty} \ab(-\frac{\tanh \zeta}{2})^n \frac{\sqrt{(2n)!}}{n!} \ket{2n}_m,
\end{equation}
and the binomial theorem~\cite{Lehmer1985}, we obtain
\begin{equation}
    \begin{aligned}
        &{}_m\bra{0}\hat{S}_m^\dagger(\zeta) e^{-i\mu_e \hat{n}_b t} \hat{S}(\zeta) \ket{0}_m \\
        &= \frac{1}{\cosh \zeta} \sum_{n=0}^{\infty} \ab(\frac{e^{-2i\mu_e t} \tanh^2 \zeta}{4} )^{n} \frac{(2n)!}{(n!)^2} \\
        &= \frac{1}{\cosh \zeta \sqrt{1 - e^{-2i\mu_e t} \tanh^2 \zeta}}.
    \end{aligned}
\end{equation}
Then, we find
\begin{equation}
    \begin{aligned}
        &\ab|_m\bra{0}\hat{S}^\dagger(\zeta) e^{-i\mu_e \hat{n}_b t} \hat{S}(\zeta) \ket{0}_m|^2 \\
        &= \frac{1}{ \ab|\cosh^2\zeta - e^{-2i\mu_e t} \sinh^2\zeta|} \\
        &= \frac{1}{\sqrt{1+ (\sin\mu_e t \sinh 2\zeta)^2}}.
    \end{aligned}
\end{equation}

Since the angle of the phase rotation arising from $\hat{\mathcal{A}}(z)$ in Eq.~\eqref{ap_eq:A_integral_form} ranges from $0$ to $\sim \mu_e/\gamma$ for the case of $\mu_e \ll \gamma$, which is the typical regime for trapped atoms, we set the ansatz for gate infidelity as
\begin{equation} \label{ap_eq:error_model}
    \begin{aligned}
        \epsilon(\Lambda) &= \epsilon^\text{recoil} + \Lambda \ab[1-\ab|_m\bra{0}\hat{S}^\dagger(\zeta) e^{-i\mu_e \hat{n}_b/\gamma} \hat{S}(\zeta) \ket{0}_m|^2 ]\\
        &= \epsilon^\text{recoil} + \Lambda \ab\{1- \frac{1}{\sqrt{1+ [\sin(\mu_e/\gamma) \sinh 2\zeta]^2}}\},
    \end{aligned}
\end{equation}
where $\epsilon^\text{recoil}$ is the infidelity at $\mu_e = \mu$, i.e., the infidelity induced only by the recoil-kick effect, and $\Lambda$ is a dimensionless free parameter.
In Fig.~\ref{fig6}, we find that $\epsilon_\text{run}(0.015)$ and $\epsilon_\text{st}(0.02)$ well reproduce the numerical results for the running- and standing-wave cavities in a wide range of $\mu_e/\mu$, respectively.
The deviation appears in the regime of $\mu_e/\mu \gtrsim 5$, since the assumption of $\mu_e \ll \gamma$ breaks down ($\gamma/\mu = 21.7$ in Fig.~\ref{fig6}).

In addition, by considering the relation
\begin{equation}
    \sinh 2\zeta = \frac{1}{2}\ab(\frac{\mu}{\mu_e} - \frac{\mu_e}{\mu}),
\end{equation}
from Eq.~\eqref{ap_eq:def_zeta} and assuming $\mu < \mu_e \ll \gamma$, we find
\begin{equation}
    \begin{aligned}
        \epsilon(\Lambda) - \epsilon^\text{recoil} &\propto 1- \frac{1}{\sqrt{1+ [\sin(\mu_e/\gamma) \sinh 2\zeta]^2}} \\
        &\simeq \frac{1}{8} \ab(\frac{\mu}{\gamma})^2 \ab(\frac{\mu_e}{\mu})^4,
    \end{aligned}
\end{equation}
which explains the scaling $(\mu_e/\mu)^4$ for the infidelity in Fig.~\ref{fig3}(b).
Note that the numerical results in Fig.~\ref{fig6} indicate that the scaling continues to hold even for $\mu_e/\mu \gtrsim 5$.
We leave the analytical modeling of such a detailed scaling relation for future work.

\bibliography{refs}
\end{document}